\documentclass[aip,reprint,english,a4paper] {revtex4-1}

\usepackage{graphicx}
\usepackage{amssymb}
\usepackage{booktabs}
\usepackage{array}
\usepackage{epstopdf}
\usepackage{amsmath}
\usepackage{mathtools}
\usepackage{amssymb}
\usepackage{upgreek}
\usepackage[normalem]{ulem}
\usepackage{color}
\usepackage{siunitx}
\DeclareSIUnit\angstrom{\text{Å}}

\usepackage{ifthen}
\usepackage[section]{placeins}
\usepackage{hyperref}
\usepackage[multidot]{grffile}
\newcolumntype{P}[1]{>{\raggedright\arraybackslash}p{#1}}

\newcommand{\dir}{../Bib}
\newcommand{\dirfig}{./}
\newcommand{\SIContactsBR} {S1}
\newcommand{\SIContactsRho} {S2}
\newcommand{\SIWaterContactsRho} {S3}
\newcommand{\SIContactsBetaAR} {S4}
\newcommand{\SILigandContacts} {S5}

\newcommand{\SIContactsActStates} {S7}
\newcommand{\SIContactsBetaARLong} {S8}

\newcommand{\SISecStructRatesBetaARLong} {S10}

\newcommand{\SIHeaterLigandRhoWater} {S1}
\newcommand{\SIHeaterLigand} {S2}
\newcommand{\SIHeaterLigandBetaARLong} {S3}

\makeatletter
\def\@firstoftwo@second#1#2{%
  \def\temp##1.##2\@nil{##2}%
   \temp#1\@nil}
\newcommand\sref[1]{%
   \expandafter\@setref\csname r@#1\endcsname\@firstoftwo@second{#1}%
}
\makeatother

\AtBeginDocument{%
	\heavyrulewidth=.08em
	\lightrulewidth=.05em
	\cmidrulewidth=.03em
	\belowrulesep=.65ex
	\belowbottomsep=0pt
	\aboverulesep=.4ex
	\abovetopsep=0pt
	\cmidrulesep=\doublerulesep
	\cmidrulekern=.5em
	\defaultaddspace=.5em
}

\begin{document}

\title{Energy Transport and its Function in Heptahelical Transmembrane Proteins}
 \author{Nadja Helmer}
 \author{Steffen Wolf}
 \author{Gerhard Stock}
 \email{stock@physik.uni-freiburg.de}
 \affiliation{Biomolecular Dynamics, Institute of Physics,
   University of Freiburg, 79104 Freiburg,~Germany}
\date{\today}

\begin{abstract}

  Photoproteins such as bacteriorhodopsin (bR) and rhodopsin (Rho)
  need to effectively dissipate photoinduced excess energy to prevent
  themselves from damage. Another well-studied
  7 transmembrane (TM) helices protein is the 
  $\beta_2$ adrenergic receptor ($\beta_2$AR), 
  a G protein-coupled receptor (GPCR) 
  for which energy dissipation paths have been
  linked with allosteric communication. To study the vibrational
  energy transport in the active and inactive states of these proteins, a
  master equation approach [J.\ Chem.\ Phys.\ {\bf 152}, 045103
  (2020)] is employed, which uses scaling rules that allow to
  calculate energy transport rates solely based on the protein
  structure.
  Despite their overall structural similarities, the three 7TM proteins reveal quite
  different strategies to redistribute excess energy. While bR quickly
  removes the energy using the TM7 helix as a 'lightning rod', Rho
  exhibits a rather poor energy dissipation, which might eventually
  require the hydrolysis of the Schiff base between the protein and
  the retinal chromophore to prevent overheating. Heating the ligand
  adrenaline of $\beta_2$AR, the resulting energy transport network of the
  protein is found to change significantly upon switching from the
  active to the inactive state. While the energy flow
  may highlight aspects of the interresidue couplings of $\beta_2$AR, it
  seems not particularly suited to explain allosteric phenomena.
  
%beyond the contact network.some connectivityin this way pathways.communication. 

\end{abstract}
\maketitle

%
%%%%%%%%%%%%%%%%%%%%%%%%%%%%%%%%%%%%%%%%%%%%%%%%%%%%%%%%%%%%%%%%%%%%%%%
%
\section*{Introduction}
Vibrational energy transport can be a critical aspect of protein
functionality. In particular, proteins that employ or generate excess energy during
their function need to quickly dissipate this energy to prevent damage
to the protein.\cite{Leitner09} For example, enzymatic
reactions may release enough heat to unfold a protein. \cite{Riedel15}
Efficient energy dissipation is also crucial for
photoproteins, which harness photonic energy via light-active
cofactors, so-called chromophores, that can generate excess
energies\cite{Champion05} even higher than \SI{2}{eV}.
Following photoexcitation, this is achieved via ultrafast internal
conversion into the electronic ground state and subsequent
redistribution of the chromophore vibrational energy into
low-frequency modes.\cite{Domcke97}
Using time-resolved vibrational spectroscopy, the resulting flow of
biomolecular energy along the protein backbone and via tertiary
contacts such as hydrogen bonds, salt bridges and polar contacts can
be monitored in space and
time.\cite{Mizutani97,Deak04,Botan07,Li14,Rubtsov19,Deniz21}
Notably, these energy transport pathways are believed to be linked to
channels of allosteric communication,\cite{Ota05, Ishikura06, Nguyen09a,
  Martinez11, Ortiz15, Weber15, Maggi18} which is of key
importance for protein signaling and regulation. \cite{Wodak19}

\begin{figure}[h!]
	\centering
	\includegraphics[width=\linewidth]{\dirfig/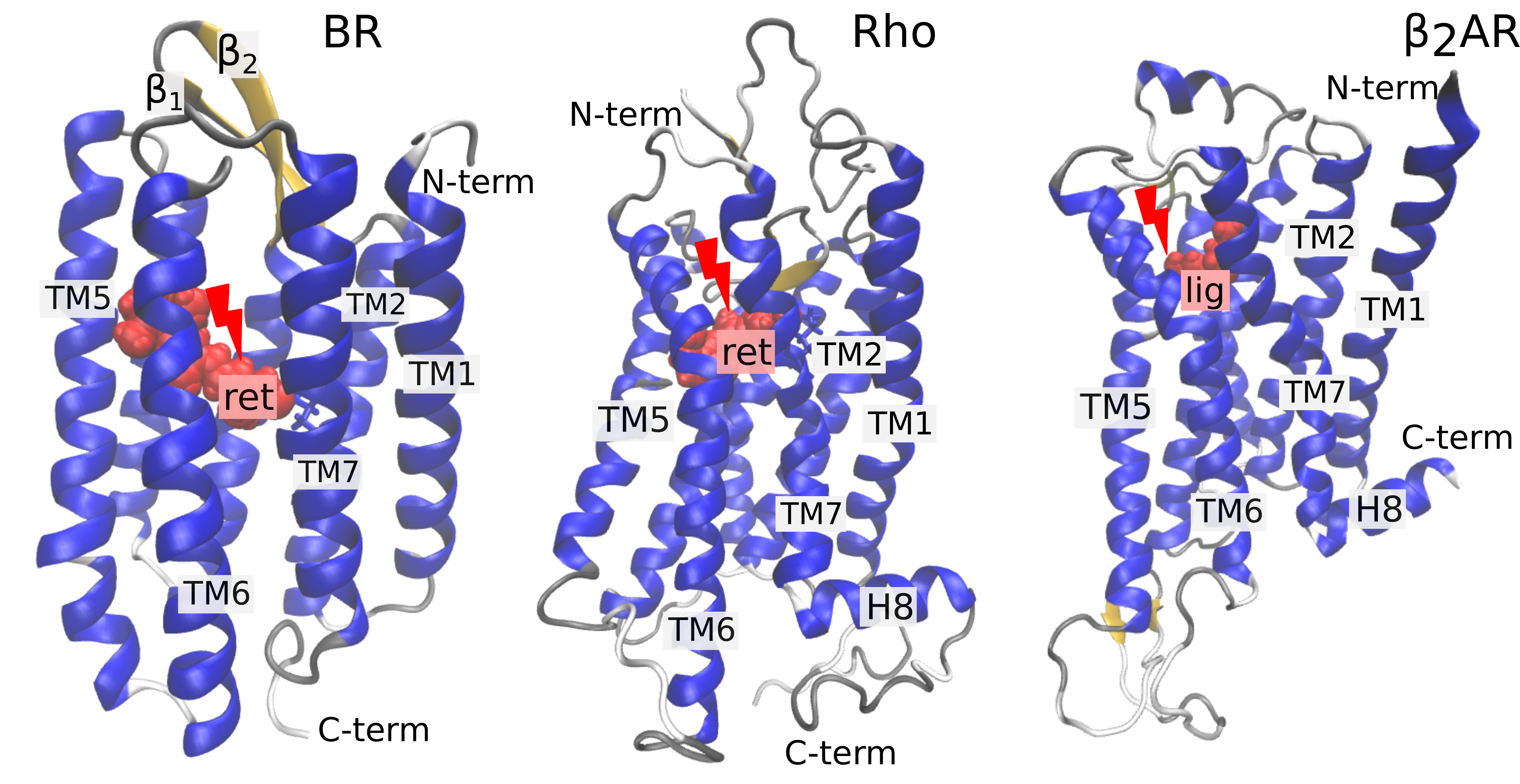}
	\caption{Structure of the three molecular systems considered
          in this work: bacteriorhodopsin, rhodopsin, and $\beta_2$
          adrenergic receptor. To initiate nonequilibrium flow of
          vibrational energy, the chromophore retinal (in bR and Rho)
          or the ligand adrenaline (in $\beta_2$AR) is heated, which
          is indicated by a red flash of lightning.}
	\label{fig:systems}
\end{figure}

Well-known examples of photoproteins are the two protein families of
microbial\cite{Briggs05,Ernst14,Rozenberg21} and
visual\cite{Smith10,Ernst14} rhodopsins. For the first family, the
prime example is bacteriorhodopsin (bR),\cite{Lanyi04} which was
discovered over 40 years ago as the only protein in the purple
membrane of \textit{Halobacterium salinarum}.\cite{Oesterhelt71} bR
acts as light driven ion pump, using the energy of green light photons
to pump protons out of a bacterial cell, creating a proton gradient
for ATP production.\cite{Oesterhelt73} The visual rhodopsin family is
best represented by rhodopsin (Rho) itself, which is the
photoreceptive protein located in the membrane of vertebrae retina rod
cells and forms a subgroup of the G protein-coupled receptor protein
(GPCR) superfamily.\cite{Fredriksson03,Lefkowitz04,Venkatakrishnan13}
As can be seen in Fig.~\ref{fig:systems}, both protein families share
seven transmembrane helices (TM1--TM7) connected by intracellular
loops (ICL) and extracellular loops (ECL) as well as a retinal
chromophore bound to the protein core and covalently linked to the
protein through a Schiff base via a lysine residue. Despite these
global similarities, the details of the protein's architecture differ
significantly, and both classes appear to have evolved
independently.\cite{Wolf15}

In both types of proteins, photon absorption induces an isomerization
of retinal,\cite{Ernst14} which drives protein conformational and
electrostatic changes necessary for the protein's function. The main
difference between bR and rhodopsin is found in the details of the
photoreaction: bR undergoes a photocycle where the retinal starts in
an \textit{all-trans} configuration and isomerizes into a
13-\textit{cis} arrangement, with the \textit{all-trans} configuration
and the initial protein conformation being restored at the end of the
photo reaction. In Rho, the retinal starts in a 11-\textit{cis}
configuration and irreversibly converts into the \textit{all-trans}
state, switching the protein from an inactive dark state to the active
Meta II state. The \textit{all-trans} retinal then needs to be
chemically cleaved from the protein and be chemically converted back
into its 11-\textit{cis} configuration by photoisomerases. Despite
these differences, $\sim$2/3 of absorbed photons result in the
isomerization in both bR and rhodopsin,\cite{Ernst14} which in turn
means that 1/3 of all absorbed photons inject their full energy into the
protein via internal conversion, as fluorescence in both proteins is
negligible.\cite{Smith85,Ernst14} Ultimately, only a part of the
absorbed photonic energy is used to perform work,\cite{Mous22} while
the rest is dissipated into the protein. Furthermore, the all-trans
retinal in Meta II is chemically cleaved from the opsin apoprotein,
unbinds from it and isomerizes back to 11-\textit{cis} retinal using
additional proteins.\cite{Piechnick12,Kiser14}  

Apart from the photoprotein bR and Rho, we are also concerned with the $\beta_2$
adrenergic receptor ($\beta_2$AR),\cite{Lefkowitz04,Kobilka11} which
represents a well-investigated GPCR that undergoes an allosteric
transition upon ligand binding.  As displayed in
Fig.~\ref{fig:systems}, $\beta_2$AR is structurally highly similar to
Rho, but is activated by the diffusible ligand adrenaline instead of a
chromophore. By heating the ligand and following the resulting energy
flow though $\beta_2$AR, one may study the change of the energy
transport network when we switch switch from the active to the inactive
state. In this way we connect to recent work \cite{Poudel21} that
linked vibrational energy transport and allostery in $\beta_2$AR.

A number of approaches have been suggested to model protein energy
flow based on molecular dynamics (MD) simulations.\cite{Nguyen03,
  Fujisaki05, Park09, Nguyen10, Soler11, Wang18, Elenewski19,
  Gulzar19, Poudel20} Moreover, several authors have proposed
network-type models,\cite{Yu03, Leitner09a, Kong09, Ortiz14,
  Leitner15, Ishikura15, Buchenberg16, Reid18} which are derived, for
instance, by computing the inter-residue energy flow to create an
energy exchange network.
To describe the energy transport in proteins between the different
residues, Buchenberg et al. have proposed a master equation
approach.\cite{Leitner15,Buchenberg16} The discovery of scaling rules
\cite{Buchenberg16,Valino-Borau20} allows for calculating the energy
transport rates solely based on the protein structure. Hence, it can
be applied on a single protein frame such as a crystal structure
without the need for extensive MD simulations.

In this work, we use the master equation approach of Ref.\
\onlinecite{Valino-Borau20} to study energy dissipation from retinal
into both bR and Rho. Additionally, we consider the ligand-induced
energy transport in $\beta_2$AR, in order to investigate a possible
connection between energy dissipation paths and allostery. For all
three proteins, high-resolution crystal structures are available in
ground (inactive) states and active
states.\cite{Okada04,Cherezov07,Choe11,
  Rasmussen11,Rasmussen11,Nango16,Weinert19} For each system, we
compute the energy transport rate matrix, which includes both
transport along the backbone and through contacts. We evaluate a
coarse-grained depiction of the rates on the level of the secondary
structures, which yields an overview over the connectivity of the
different proteins. By heating the ligand or retinal, respectively, we
visualize and discuss the resulting energy flow in some detail.

%
%%%%%%%%%%%%%%%%%%%%%%%%%%%%%%%%%%%%%%%%%%%%%%%%%%%%%%%%%%%%%%%%%%%%%%%
%
\section*{Theory and Methods}
\subsection*{Master Equation}
\vspace{-.4cm}
Starting from a discretized diffusion equation, Buchenberg et
al. \cite{Buchenberg16} derived the energy flow master equation 
\begin{align}
\frac{dE_j(t)}{dt} = \sum_i \left[k_{ij}E_i(t) - k_{ji}E_j(t)\right], \label{Master}
\end{align}
where $E_{j}$ describes the kinetic energy of residue $j$ and $k_{ij}$
denotes the transport rate from residue $i$ to $j$. The corresponding
energy transport times $\tau_{ij}$ are obtained via
$\tau_{ij} = 1/k_{ij}$.  The rate matrix ${k_{ij}}$ contains the full
information about the energy redistribution, and can be employed to
predict the energy evolution of the system. Moreover, one can consider
the solvent as an additional state with solvent rates $k_{ps}$ and
$k_{sp}$ from the protein into the solvent and back. \cite{Park09}
However, we do not include a solvent rate in this work, as the major
protein surface is found at the membrane interface, and the protein-water
interface above and below the
membrane only leads to a non-specific removal of the excess 
heat. \cite{Valino-Borau20,Deniz21} 
Protein-membrane lipid contacts are mostly hydrophobic, and as was found earlier,
\cite{Buchenberg16, Poudel20} the energy transfer through such
contacts can be safely neglected.

\subsection*{Scaling rules and Identification of Contacts}
\vspace{-.4cm}

%Applying the master equation~(\ref{Master}), Valino Borau et
%al. \cite{Valino-Borau20} recently presented a new parametization
%%based on extensive nonequilibrium MD simulations. 
The scaling rules of biomolecular energy flow allow to describe the
energy transport in terms of only a few parameters, yielding a simple
and transferable model. The scaling rule for backbone transport yields
the rates \cite{Buchenberg16}
\begin{align}
k_{ij}^B = \frac{D_\mathrm{B}}{\langle x_{ij}\rangle^2} \sqrt{\frac{f_j}{f_i}}, \label{BB_rate}
\end{align}
and for contact transport we obtain \cite{Valino-Borau20}
\begin{align}
k_{ij}^C = \frac{D_\mathrm{C}}{\langle q_{ij}\rangle^2} \sqrt{\frac{f_j}{f_i}}. \label{contact_rate}
\end{align}
Both rules follow from the diffusion equation with $D_\mathrm{B}$ and
$D_\mathrm{C}$ being diffusion constants and $f_i$ denoting the number
of the degrees of freedom of residue $i$.  In case of the backbone
transport, we consider the average square distance
$\langle x_{ij}\rangle^2$ along covalent bonds between all atoms of
residues $i$ and $j$. In contrast to that, the contact transport only
takes the atoms of selected contacts into account (see below) and is
inversely proportional to the square mean distance
$\langle q_{ij}\rangle^2$ between those atoms. The diffusion constants
were determined as $D_\mathrm{B} = \SI{1.1}{nm^2 ps^{-1}}$ and
$D_\mathrm{C} = \SI{2.1 e-3}{nm^2 ps^{-1}}$ in Ref.\
\onlinecite{Valino-Borau20} and serve as global constants, while the
individual bond strength is reflected by $\langle x_{ij}\rangle^2$ and
$\langle q_{ij}\rangle^2$, respectively. The application of these
scaling rules allow us to predict the energy transport directly from
the connectivity within a protein structure, providing the general
means to model the energy flow in various proteins using solely a
crystal structure.

Since nonpolar contacts were shown to be negligible for energy
transport,\cite{Buchenberg16, Poudel20} they are omitted in this
study. Moreover, as the residues that form strong polar contacts
additionally also form hydrogen bonds (but with a smaller distance),
it is sufficient to focus on hydrogen bonds for energy transport
as a first approximation.  Here we only include hydrogen bonds, where
the distance between hydrogen atom and acceptor is smaller than
\SI{2.8}{\AA} and the angle $\theta$ between donor-acceptor-hydrogen
is $\theta \leq 30^{\circ}$. 
Concerning ionic interactions in the considered proteins, charged
residues interact with each other via salt bridges, i.e., combined
ionic-hydrogen bond interactions. Hence, this type of interaction is
already approximately taken care of via the hydrogen bond
analysis. We furthermore consider ionic interactions over
distances larger than 4.5~\AA\ to be damped out by other residues found
between them, and thus so weak that their contribution to energy
transfer is negligible.

>From these contacts, we compute the energy transport rates on the
level of residues, as those are the states used in the master
equation. To obtain a broader picture, we also consider the energy
transport on the level of the protein's secondary structures, where
all rates between the residues of two secondary structures are
combined into one rate. This method allows us to obtain a better
overview of the connections between the different protein
structures. The secondary structure rates can then be visualized in
the protein structure to grasp the protein's connectivity at one glance.

\subsection*{Quantum Corrections}
\vspace{-.4cm} 

The above scaling rules were parameterized from classical
nonequilibrium MD simulations.\cite{Valino-Borau20} However, classical
mechanics approximate the quantum mechanics of anharmonic oscillators
only well on short time scales.\cite{Egorov99, Schade09, Stock09} In
order to reproduce the correct energy transport time scales, it is
thus necessary to introduce a quantum correction to our classical
approach. As a simple approximation, it may be sufficient to rescale
the diffusion coefficients $D_\mathrm{B}$ and $D_\mathrm{C}$ obtained
from the classical MD simulations by a common quantum correction
factor $Q$. By comparison of simulated and experimentally measured
data for various test proteins, this factor was determined as $Q=3.1$
by Deniz et al.\cite{Deniz21} As in this work we no not include the
dissipation into the solvent water (which typically needs no
correction \cite{Park09}), the quantum correction is applied to all
considered transport channels and therefore only serves as a general
scaling factor to obtain the correct the time scales.

\subsection*{Structure Preparation}
\vspace{-.4cm} In order to compute the contact rates, we first have to
identify the contacts that the protein forms.  We apply our method to
crystal structures, which were in part slightly modified.  For bR, we
base our calculations on the recently resolved set of crystal
structures by Weinert et al.\cite{Weinert19} We consider the inactive
"dark" state (PDB 6RQP) with an \textit{all-trans} retinal and the
active state with 13-\textit{cis} retinal (PDB 6RPH), which represents
the subsequent N- and O-intermediates in the photocycle.\cite{Lanyi04}
In this case, the crystal structures already include hydrogen atoms,
which we use for further contact analysis.
For Rho, we consider the inactive dark state
(PDB 1U19), where the retinal is bound in the \textit{cis} state, and
the Meta II state (PDB 3PQR), with the isomerized retinal in the
\textit{all-trans} state still bound to the protein.  
The results for $\beta_2$AR relies on crystal structures of the
inactive \cite{Cherezov07} (PDB 2RH1) and the active
state\cite{Rasmussen11} (PDB 3P0G) with the docked native ligand
adrenaline. To verify that docking a ligand is a viable approach for
studying energy transmission in proteins, we additionally considered a
crystal structure of an active $\beta_2$AR-adrenaline
complex\cite{Ring13} (PDB 4LDO). Missing hydrogen atoms are added to
protein crystal structures with PROPKA3\cite{Olsson11,Sondergaard11}
for a pH of 7 using the PDB2PQR web server.\cite{Dolinsky04,Jurrus18} The missing
intracellular loop 3 in the $\beta_2$AR was recovered via the
SWISS-MODEL web server,\cite{Waterhouse18} and the native ligand
adrenaline was added via docking using Autodock Vina.\cite{Trott10}
T4 lysozyme fusion proteins and active state stabilizing
peptides or proteins were removed. To handle the structural protein information,
we use \textit{MDAnalysis}.\cite{MDAnalysis1, MDAnalysis2}

For MD simulations of $\beta_2$AR, the same structure preparation 
protocol was employed. In the case of the active receptor, 
we employed PDB structure 4LDO with the nanobody present to stabilize
the receptor. The complex was embedded into a POPC lipid bilayer
surrounded by water using INFLATEGRO2. \cite{Schmidt12}

\subsection{Molecular Dynamics Simulations}

Atomic interactions were described by the Amber99SB* force
field\cite{Hornak06,Best09} combined with the Berger lipid 
force field,\cite{Berger97,Cordomi12} and the TIP3P water
model.\cite{Jorgensen83} Adrenaline parameters were generated
using Antechamber\cite{Wang06} and Acpype\cite{Sousada12} with atomic
parameters derived from GAFF parameters\cite{Wang04} and AM1-BCC
atomic charges. \cite{Jakalian00,Jakalian02}

Simulations were carried out using Gromacs
v2018 (Ref.~\onlinecite{Abraham15}) in a CPU/GPU hybrid
implementation. Van der Waals interactions were calculated with
a cut-off of 1~nm, electrostatic interactions using the particle mesh
Ewald method\cite{Darden93} with a minimal real-space cut-off of
1~nm. All covalent bonds with hydrogen atoms were constrained using
LINCS.\cite{Hess97} After an initial steepest descent minimization
with positional restraints of protein and ligand heavy atoms, an
initial 10~ns equilibration MD simulation in the NPT ensemble was
performed with a 2~fs time step and positional restraints of protein
and ligand heavy atoms. A temperature of 300~K was kept constant
using the Nos\'e-Hoover thermostat\cite{Nose84,Hoover85} (coupling time
constant of 0.2~ps), the pressure was kept constant at 1~bar using the
Berendsen barostat\cite{Berendsen84} (coupling time constant of
0.5~ps) with semi-isotropic pressure coupling, followed by a 
second steepest descent minimisation without
restraints and a short 0.1~ns equilibration MD simulation in the NPT
ensemble.
Then, 120~ns of free MD simulations were carried out, switching the barostat
to the Parrinello-Rahman barostat.\cite{Parrinello81} Snapshots were collected 
each 20~ns, resulting in a total of 7 structures.

\subsection*{Heating Process and Visualization}
\vspace{-.4cm}

Performing extensive nonequilibrium MD simulations of the energy flow
in proteins, Gulzar et al.\cite{Gulzar19} approximated the initial
photoexcitation by an instantaneous temperature jump, where the
resulting excess energy $k_{\rm B}\Delta T$ of the chromophore is
chosen to match the $S_0\!\rightarrow \! S_1$ excitation energy of
$\approx$ 2~eV, resulting in $\Delta T \approx 600$ K. In the master
equation simulations of the energy flow, we use this temperature
change as initial condition at time $t=0$ to define the initial energy
of the chosen heater system (i.e., retinal for bR and Rho and the
ligand for $\beta_2$AR). We note that this choice does only affect the
observed temperature scale but not the energy transfer dynamics of the
master equation.

Solving the master equation \eqref{Master} iteratively, we obtain
the time evolution of the energy of every residue of the protein. To
illustrate the energy transport in a more visual manner, we utilize
these temperature curves to color the residues for each time step
using \textit{VMD}.\cite{Humphrey96} This visualization allows us to
watch the energy distribution in form of a movie and choose suitable
time frames for representation. In the end, the energy distributes
until equilibrium is reached and the equipartition theorem is
fulfilled.

\section*{Results}
\subsection*{Bacteriorhodopsin}
\vspace{-.4cm} 
Starting with bR, we find 34 contacts for the inactive
state and 32 for the active state, which we list in
Tab.~\SIContactsBR.  To analyze how the system deals with the energy
of the absorbed photons, we heat the retinal chromophore, which is
covalently bound to Lys216 in TM7.

\begin{table}
  \caption{Secondary structure rates of bR for the active and inactive
    state. The rates are given in $\si{ps^{-1}}$.} 
	\label{tab:sec_rates_BR}
	\centering
	\begin{tabular}{llcc}
		\toprule
		\multicolumn{2}{c}{Sec. Struct.} & \multicolumn{2}{c}{Rates}\\
		\midrule
		Sec 1 &    Sec 2 &  Inactive &   Active \\ 
		\cmidrule(lr){1-2}  \cmidrule(lr){3-4} 
		N-term               &  TM2               &    0.26 &    0.11  \\
		TM1                   &  $\beta_2$         &    0.26 &    0.16  \\
		TM1-TM2               &  C-term            &    0.09 &          \\
		TM2                   &  TM3               &    0.16 &          \\
		TM2                   &  TM7               &    0.16 &    0.15  \\
		TM2-$\beta_1$         &  TM3               &    0.14 &    0.10  \\
		$\beta_1$                   &  $\beta_2$         &    0.75 &    0.46  \\
		TM3                   &  TM7               &         &    0.11  \\
		TM3-TM4               &  TM5               &    0.21 &          \\
		TM4                   &  TM5               &    0.34 &    0.34  \\
		TM5                   &  TM6               &    0.12 &    0.08  \\
		TM5                   &  TM6-TM7           &    0.24 &    0.17  \\
		TM5-TM6               &  C-term            &    0.25 &          \\
		TM6                   &  TM7               &    0.15 &    0.10  \\
		\bottomrule
	\end{tabular}
\end{table}

\begin{figure}
	\centering
	\includegraphics[width=.8\linewidth]{\dirfig/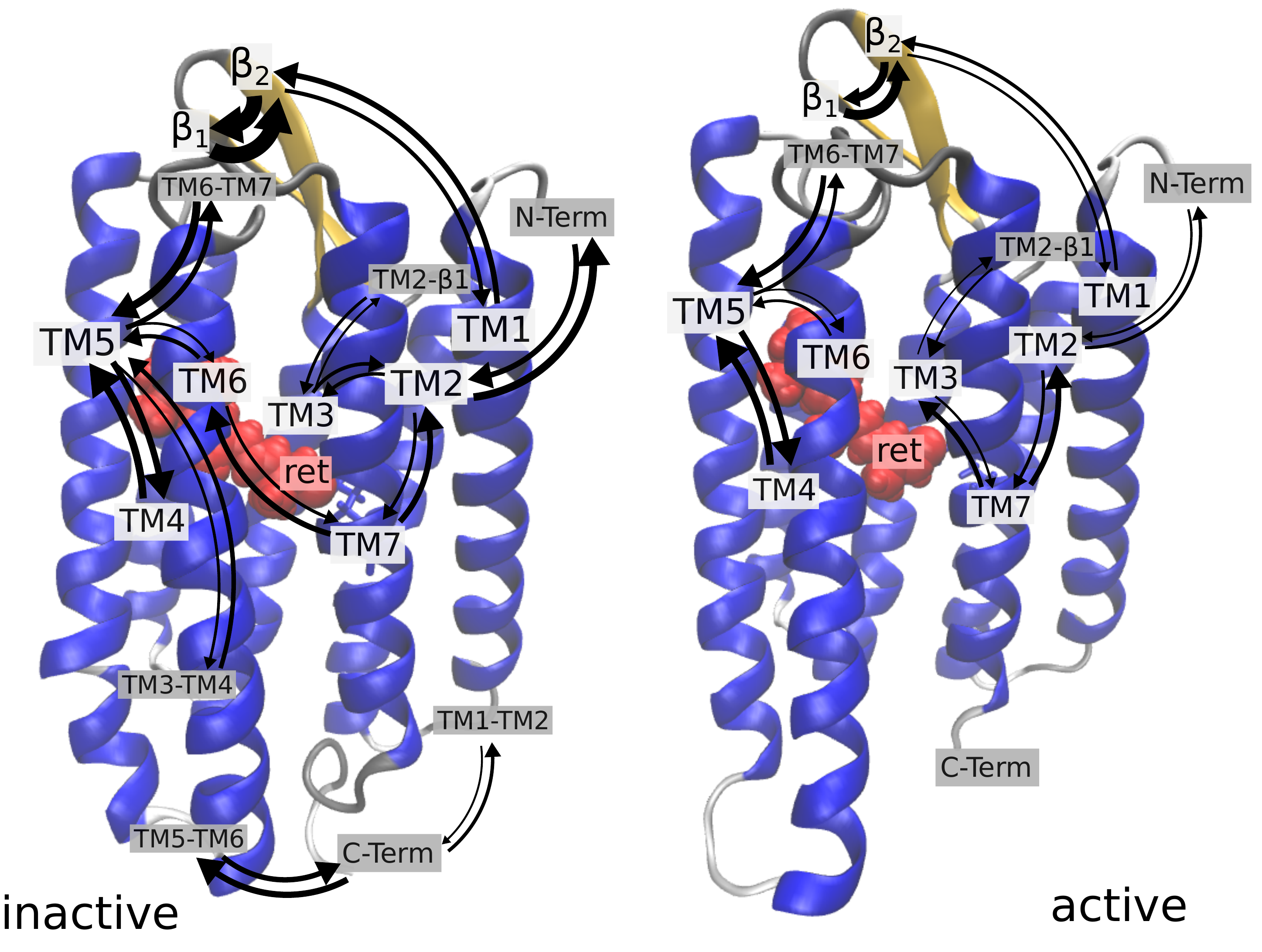}
	\caption{Illustration of the secondary structure rates of bR
          as black arrows within a snapshot of the protein. The line
          width of the arrows are proportional to the rates. For each
          structure pair, the rates are shown in both directions.}
	\label{fig:sec_rates_BR}
\end{figure}

After computing all contact and backbone rates, we group them
according to the involved secondary structures. The resulting
secondary structure contact rates can be found in
Tab.~\ref{tab:sec_rates_BR} and Fig.~\ref{fig:sec_rates_BR}. The rates
provide a rough picture of the connectivity between the different
parts of the protein. The two states mostly agree in connectivity and
absolute value of the rates. Though the largest structural difference
between both states is the outward motion of helices E and
F\cite{Weinert19} (here TM5 and TM6), the largest differences in the
rates are found for helix G (here TM7). This is most likely due to the
weakening of the Tyr185-Asp212 hydrogen bond, which elongates from
2.6~\AA\ in the dark state to 3.1~\AA\ in the active state. In
general, there seem to be hardly any noticeable changes in the energy
transport upon the activation of the protein. The main reason for this
is that most contacts within the core of bR are hydrophobic and only a
handful of contacts break during the photocycle. Furthermore, TM7
exhibits only two polar contacts via Asp204 and Asp212, hence energy
transport via the backbone chain should dominate in both investigated
structures. 

\begin{figure}[htpb]
	\centering
	\includegraphics[width=0.9\linewidth]{\dirfig/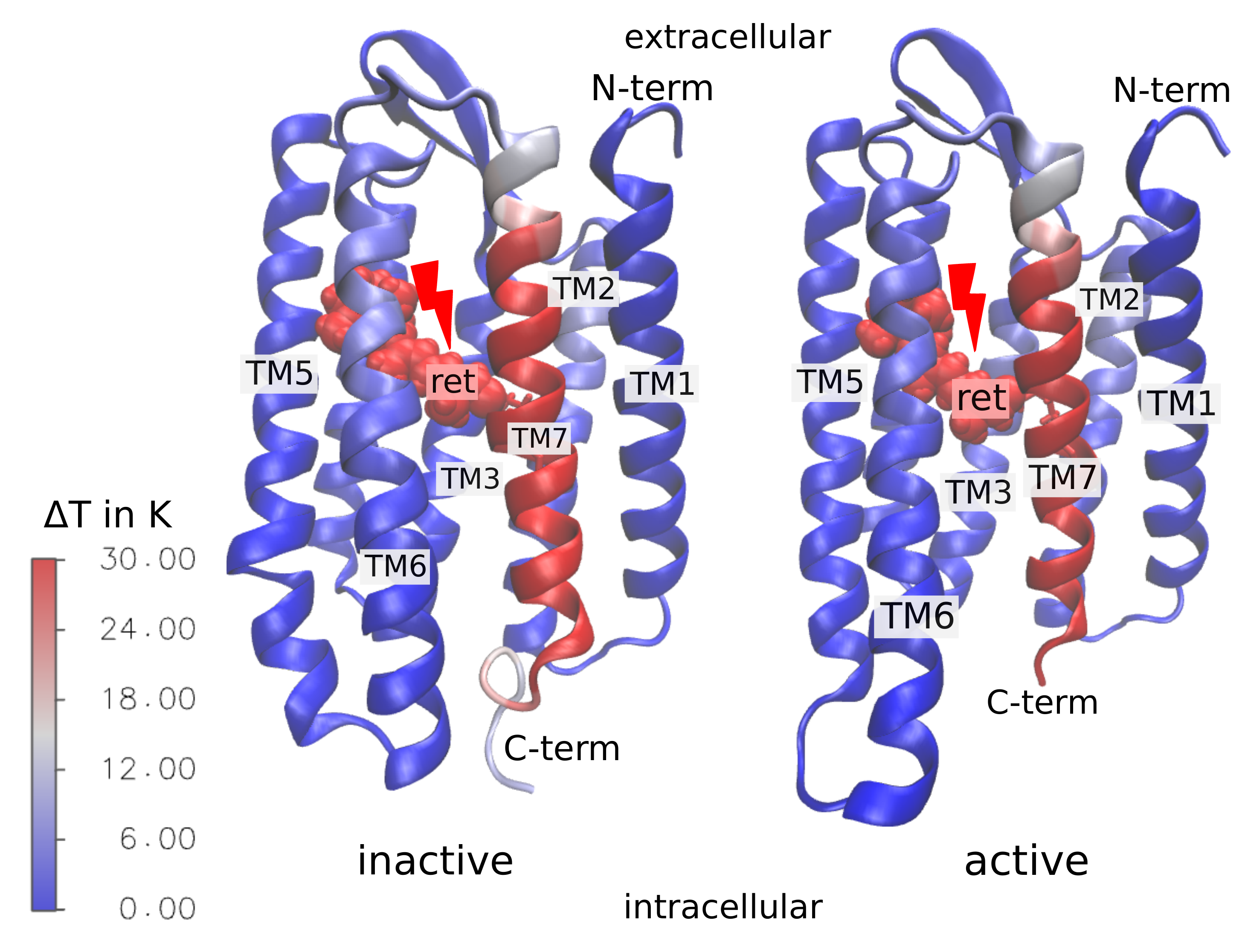}
	\caption{BR at $t = \SI{10}{ps}$ after heating the
          retinal. The left side shows the protein in the inactive
          state, the right side in the active state. Colors of the
          residues indicate the residue's temperature. } 
	\label{fig:heat_10ps_BR}
\end{figure}

Figure \ref{fig:heat_10ps_BR} shows both states of the system at
\SI{10}{ps} after heating the ligand impulsively (see Methods). As
discussed above, the energy transport in the two states is very
similar in behavior. We can observe that the energy flows through the
covalent bond into TM7 and spreads in both directions from
there. Additionally, TM3 and TM2 obtain small amounts of energy
through contact transport. However, as presumed, backbone transport
strongly dominates. We note that the C-terminus of TM7 quickly
receives a large amount of kinetic energy. In bR crystal structures,
the 16 amino acid C-terminus is not resolved, and forms a random
coil\cite{Wallace84} that is embedded in the intracellular water
bulk. Its removal does not affect the protein's folding and function
as a proton pump,\cite{Liao84} but has been linked to protein
stability.\cite{Turner09} Naturally, these amino acids will exhibit
contacts with water molecules, and form an efficient position for
dissipation of excess energy into the solvent. Figuratively speaking,
the TM7 helix seems to act like a ''lightning rod'' providing a direct
energy transfer path to the C-terminus, which forms a ''grounding
rod'' to effectively dissipate excess energy into the solvent.

\subsection*{Rhodopsin}
\vspace{-.4cm}

\begin{table}[h!]
	\caption{Rates between the different secondary structures in
          Rho for both states. The rates are given in $\si{ps^{-1}}$.} 
	\label{tab:sec_rates_Rho}
	\centering
	\begin{tabular}{llrr}
		\toprule
		\multicolumn{2}{c}{Sec. Struct.} & \multicolumn{2}{c}{Rates}\\
		\midrule
		Sec 1 &   Sec 2 & Inactive  &  Active\\
		\cmidrule(lr){1-2}  \cmidrule(lr){3-4} 
		N-Term &  TM2    &         &    0.10\\
		N-Term &  ECL1   &    0.30 &    0.11\\
		N-Term &  ECL2   &    0.21 &        \\
		N-Term &  ECL3   &    0.16 &        \\
		TM1    &  TM2    &    0.16 &    0.25\\
		TM1    &  TM7    &    0.41 &    0.15\\
		TM2    &  TM4    &    0.22 &    0.13\\
		TM2    &  ECL2   &         &    0.27\\
		TM3    &  ECL2   &    0.14 &    0.15\\
		TM3    &  TM5    &    0.15 &        \\
		TM3    &  TM6    &    0.18 &        \\
		TM3    &  TM7    &    0.11 &        \\
		TM4    &  TM5    &    0.35 &    0.49\\
		ECL2   &  TM6    &    0.17 &    0.14\\
		TM5    &  TM6    &         &    0.39\\
		TM6    &  H8     &    0.10 &        \\
		TM7    &  H8     &    0.17 &        \\
        \bottomrule
	\end{tabular}
\end{table}

\begin{figure}[h!]
	\centering
	\includegraphics[width=0.9\linewidth]{\dirfig/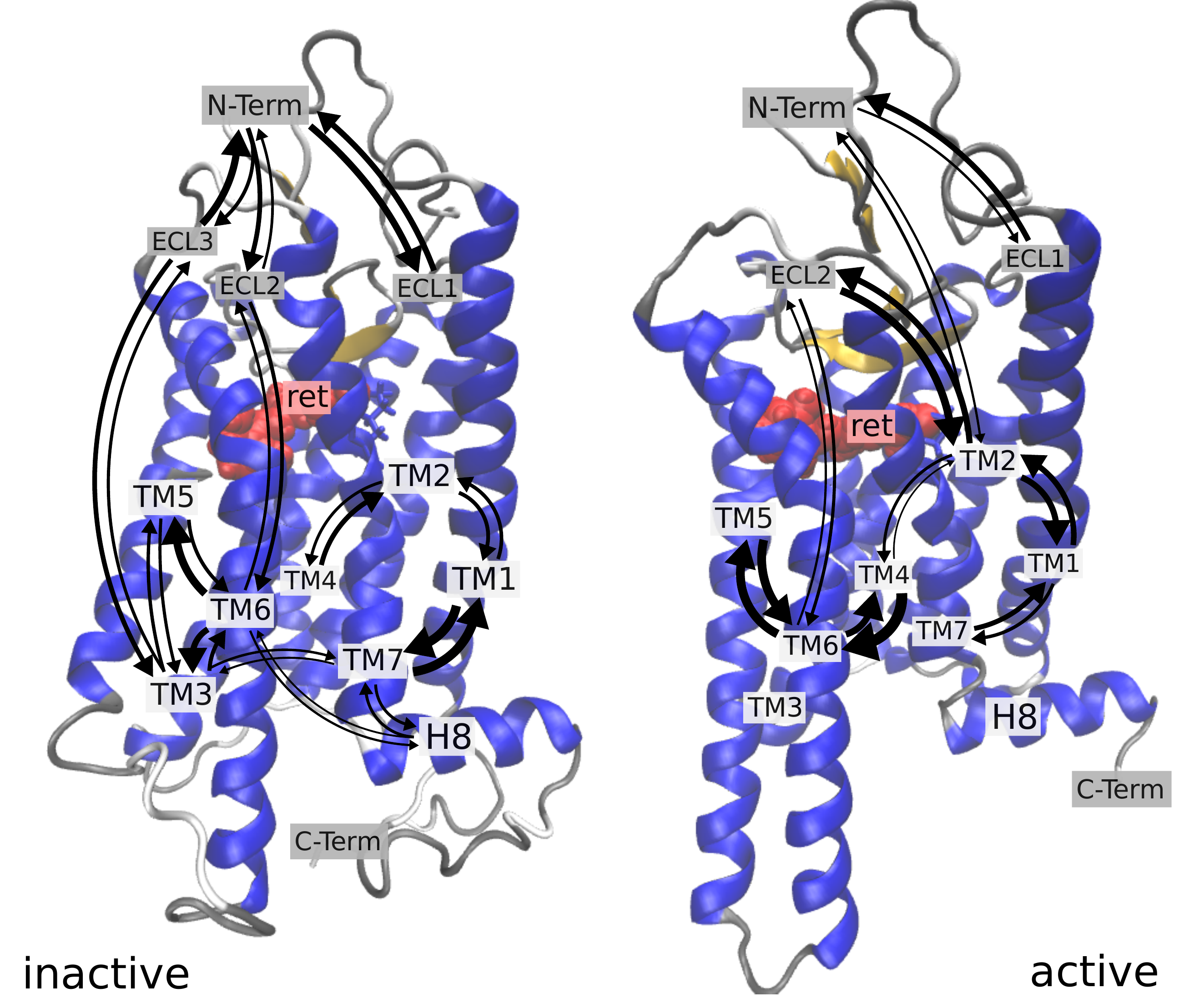}
	\caption{Visualization of the secondary structures rates of
          Rho with black arrows indicating the energy transport
          between the structures. The arrow line width are
          proportional to the secondary structure rates.} 
	\label{fig:sec_rates_Rho}
\end{figure}

We now turn to the analysis of energy transfer in Rho. We find 51
contacts between different residues for the dark state and 50 for the
Meta II state, which are listed in Tab.~\SIContactsRho. The
secondary structure rates can be found in Tab.~\ref{tab:sec_rates_Rho}
with an illustration in Fig.~\ref{fig:sec_rates_Rho}.  The inactive
state contains slightly more internal connections and thus seems to be
better at dissipating energy. Especially the two helices TM1 and TM7
are strongly connected with a large contact rate. In contrast to this,
the active state appears to be more isolated, as many of the contacts
break through changes in the structure upon activation.

\begin{figure}
	\centering
	\includegraphics[width=0.9\linewidth]{\dirfig/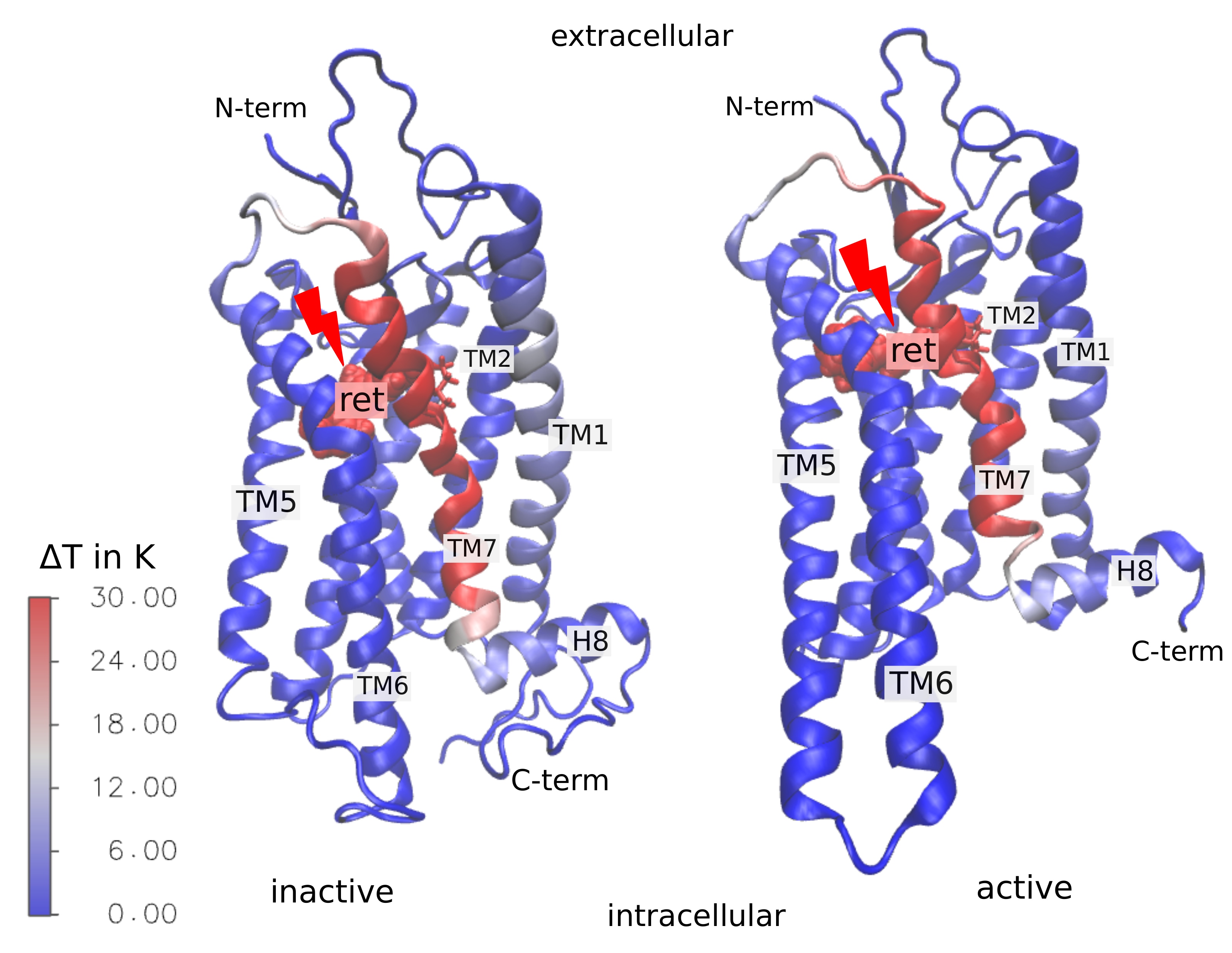}
	\caption{Rho shown in both states at $t = \SI{10}{ps}$ after
          heating the retinal. The heated retinal is indicated by a
          red lightning bolt. The coloration shows the local
          temperatures of residues.} 
	\label{fig:heat_10ps_Rho}
\end{figure}

As before we heat the retinal and observe how the energy distributes
from there.  Fig.~\ref{fig:heat_10ps_Rho} shows the two states of
rhodopsin \SI{10}{ps} after the heating process. The energy flows into
TM7 through the covalent bond and quickly spreads in both directions
along the backbone. However, the energy does not reach further than
the ends of this helix. In the inactive state, we can additionally
observe contact transport from TM7 into TM1 and TM3, even though the
energy flow towards TM3 seems negligible. The main path for this
energy transfer is the salt bridge between the protonated Schiff base
of the retinal and Glu113, which is a major structural feature to
stabilize the dark state.\cite{Kim04} In the active state, on the
other hand, no noticeable contact transport can be observed. This
agrees with the lower connectivity we deduced from the rates.

In comparison with bR, we note that the excess energy is not quickly
transferred to solvent-exposed unstructured domains, but remain in the
core of the transmembrane helix bundle. In Rho, the C-terminus serves
as important structural feature for signal transduction\cite{Tsai19}
and thus cannot serve as ''grounding rod'' as in bR. Instead, it
appears that excess energy is distributed into the whole protein, and
then slowly released from there to its environment. We notice that the
salt bridge between the Schiff base and Glu113 ruptures only after
retinal isomerization,\cite{Ernst14} i.e., after the decay back into
the retinal electronic ground state. Therefore, this salt bridge can
indeed serve as energy dissipation channel. We finally note that the
\textit{all-trans} retinal in the active state still can absorb
photons, which do not contribute to protein function
anymore. Hydrolysis of the Schiff base on the formation of opsin may
therefore not only serve for the reloading with 11-\textit{cis}
retinal, but also for the protection of rhodopsin from excess photon
energy.

It has been shown that in both bR and Rho, protein-internal water
molecules are essential for protein function and connect distant amino
acid side chains to each other via hydrogen bonds.\cite{Gerwert14}
We therefore tested for the example of Rho how much the treatment of water
molecules observed in crystal structures as contact residues would
contribute to energy transfer from the retinal to the protein. As
shown in Tab.~\SIWaterContactsRho, we found a significant number of
contacts with internal water molecules (37 in the inactive and 17 in
the active state), however, the respective contact rates are mostly
small. Hence, these water molecules represent dead ends for the energy
transfer and do not significantly change the time-dependent energy
distribution (Fig.~\SIHeaterLigandRhoWater). Protein-internal water
molecules therefore are negligible for the overall energy transport in
Rho.

\vspace{-.6cm}
\subsection*{$\beta2$ adrenergic receptor}
\vspace{-.4cm}

\begin{figure}[h!]
	\centering
	\includegraphics[width=.8\linewidth]{\dirfig/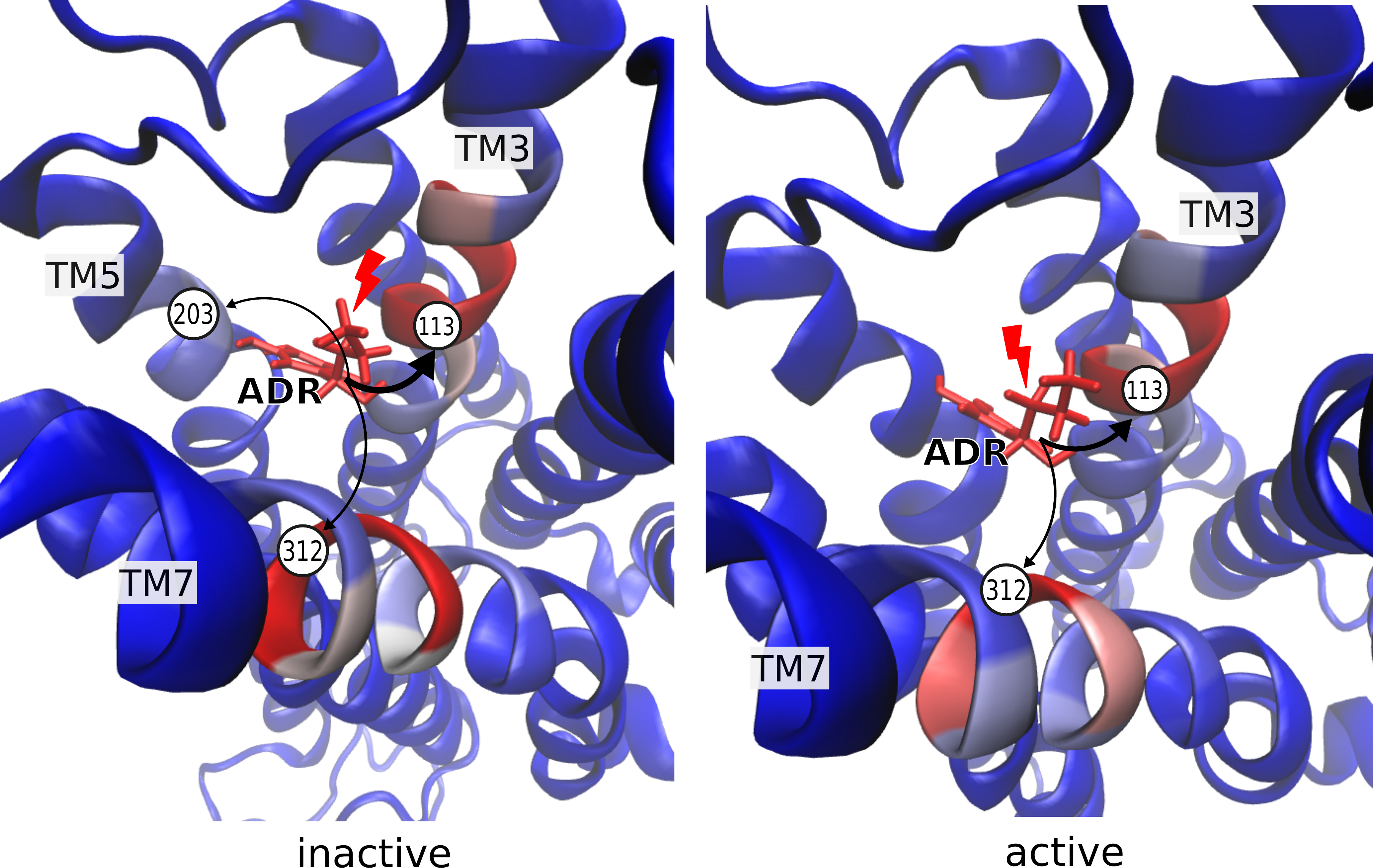}
	\caption{Illustration of the contacts rates of $\beta_2$AR for
          the inactive state (left) and the active state (right).} 
	\label{fig:B2AR_ligand_contacts}
\end{figure}

Finally, we explore if the lessons learned from energy transfer in bR
and Rho can be transferred to the energy transport in $\beta_2$AR.
%and its possible connection to allosteric communication.
%
As stated before, $\beta_2$AR binds a ligand by forming polar and
nonpolar contacts. In contrast to the other two systems, where the
retinal was covalently bound to the protein and the energy entered the
protein through backbone transport along those bonds, here the energy
flows into the protein through contact transport. Moreover, both
states form at least 2 ligand contacts and the energy consequently
already reaches different parts of the protein directly after the
heating. The contacts formed with the ligand are shown in
Fig.~\ref{fig:B2AR_ligand_contacts} and listed in
Tab.~\ref{tab:B2AR_ligand_contacts}. In both states, the ligand binds
to TM3 via a salt bridge / hydrogen bond combination with Asp113
  and to TM7 via a bidentate hydrogen bond with Asn312. Additionally,
the inactive state forms another contact through TM5 via Ser203,
however, only with a small rate.

\begin{table}[h!]
	\caption{Rates of the contacts formed between the ligand
          adrenaline and $\beta_2$AR, given in $\si{ps^{-1}}$. ''Adr''
          denotes adrenaline. } 
	\label{tab:B2AR_ligand_contacts}
	\centering
	\begin{tabular}[t]{cccc}
		\toprule
		\multicolumn{2}{c}{Residues} & \multicolumn{2}{c}{Rates}\\ 
		\midrule
		%Res 1 & Res 2 & Inactive & Active \\ \cmidrule(lr){1-2}  \cmidrule(lr){3-4}  
		%Asp113 &  Adr343  &    0.71 &     0.57 \\	
		%Ser203 &  Adr343  &    0.29 &           \\
		%Asn312 &  Adr343  &    0.27 &      0.27 \\
		%\bottomrule
		Res 1 & Res 2 & Inactive & Active \\ \cmidrule(lr){1-2}  \cmidrule(lr){3-4}  
		Adr  & Asp113 &    0.34 &     0.27 \\	
		Adr  & Ser203 &    0.12 &           \\
		Adr  & Asn312 &    0.14 &      0.14 \\
		\bottomrule
	\end{tabular}
\end{table}

\begin{figure}[h!]
	\centering
	\includegraphics[width=0.9\linewidth]{\dirfig/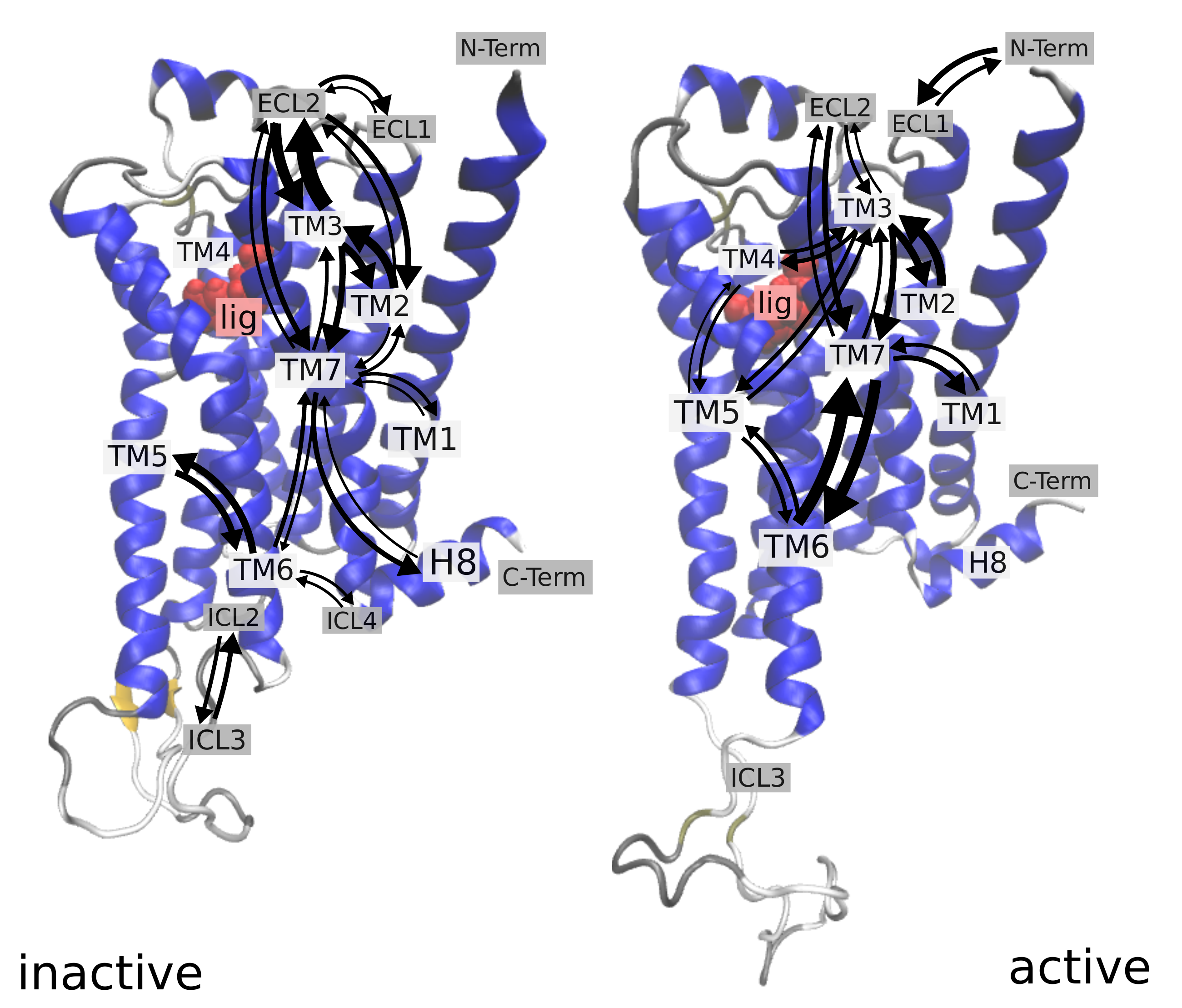}
	\caption{The secondary structure contact rates of $\beta_2$AR
          visualized in a snapshot of the protein through black arrows
          with an arrow line width is proportional to the rates.} 
	\label{fig:sec_rates_B2AR}
\end{figure}

>From the contact analysis of $\beta_2$AR we find 61 contact rates in
the inactive state and 50 in the active. This is a larger difference
than in the two proteins before and already indicates that the protein
has a higher connectivity before the activation. The contact rates are
listed in Tab.~\SIContactsBetaAR.
Considering the secondary structure rates displayed
Fig.~\ref{fig:sec_rates_B2AR}, the energy transport seems to
completely change between the two states. A list of the rates can be
found in in Tab.~\ref{tab:sec_rates_B2AR}. Many of those changes
involve TM5 and TM6, which are known for their outward movements when
the protein transitions into the active state.\cite{Rasmussen11}  

\begin{table}[h!]
	\caption{Secondary structures in $\beta_2$AR shown for both
          states. The rates are given in $\si{ps^{-1}}$.} 
	\label{tab:sec_rates_B2AR}
	\centering
	\begin{tabular}{llrr}
		\toprule
		\multicolumn{2}{c}{Sec. Struct.} & \multicolumn{2}{c}{Rates}\\ 
		\midrule
		Sec 1 &           Sec 2 &  Inactive   &  Active \\
		\cmidrule(lr){1-2}  \cmidrule(lr){3-4} 
		N-Term &  ECL1   &         &     0.25 \\
		TM1    &  TM7    &    0.11 &     0.17 \\
		TM2    &  TM3    &    0.39 &     0.40 \\
		TM2    &  ECL2   &    0.15 &          \\
		TM2    &  TM7    &    0.12 &          \\
		ECL1   &  ECL2   &    0.11 &          \\
		TM3    &  TM4    &         &     0.19 \\
		TM3    &  ECL2   &    0.71 &     0.11 \\
		TM3    &  TM5    &         &     0.17 \\
		TM3    &  TM7    &    0.18 &     0.28 \\
		ICL2   &  ICL3   &    0.18 &          \\
		TM4    &  TM5    &         &     0.14 \\
		ECL2   &  TM7    &    0.24 &     0.26 \\
		TM5    &  TM6    &    0.29 &     0.19 \\
		TM6    &  TM7    &    0.19 &     0.54 \\
		TM6    &  ICL4   &    0.13 &          \\
		TM7    &  H8     &    0.23 &          \\
		\bottomrule
	\end{tabular}
\end{table} 

\begin{figure}[htpb]
	\centering
	\includegraphics[width=0.9\linewidth]{\dirfig/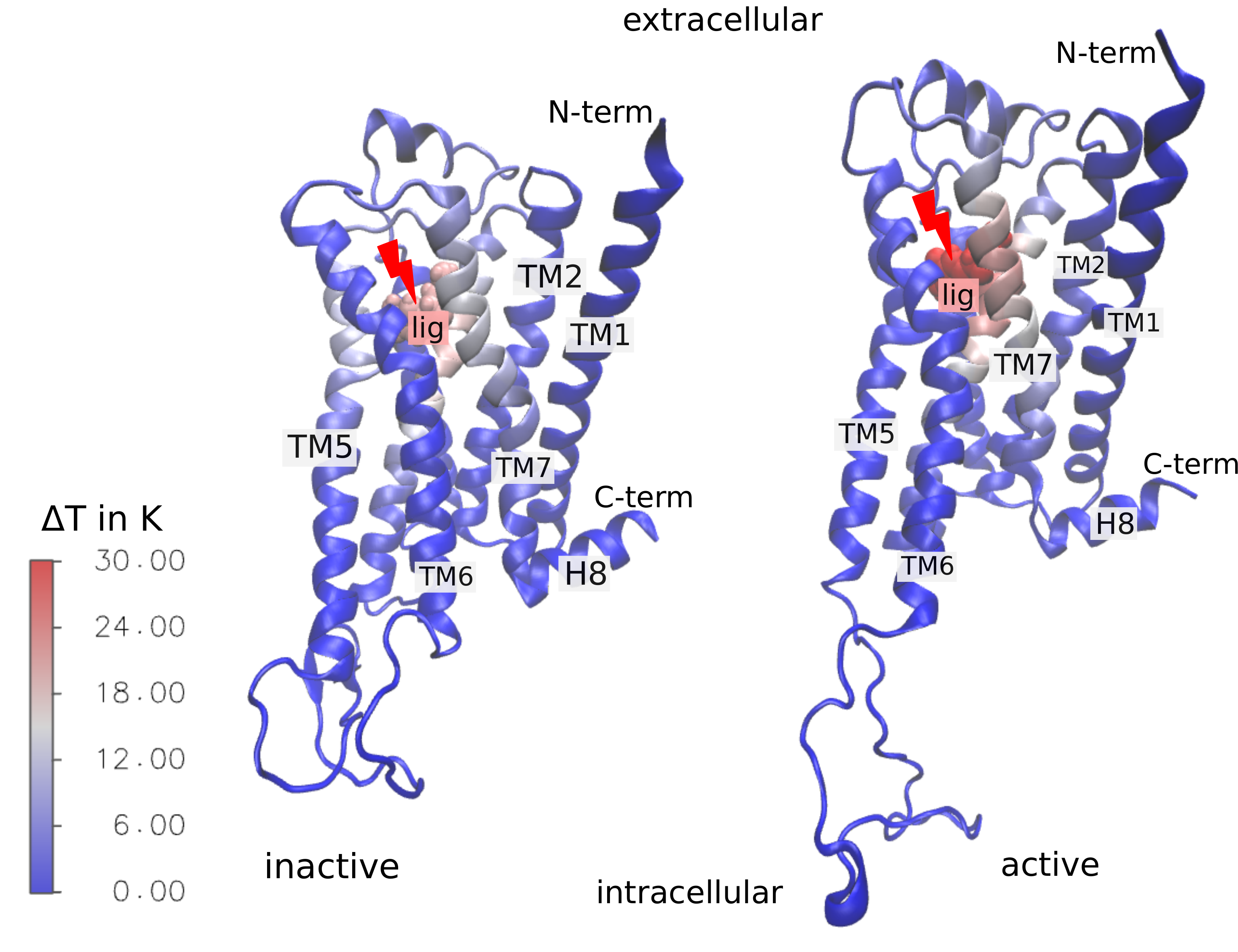}
	\caption{$\beta_2$AR shown \SI{10}{ps} after heating the
          ligand. The heated ligand adrenaline is indicated by a red
          lightning bolt. The coloration shows the local temperatures
          of residues.} 
	\label{fig:heat_10ps_B2AR}
\end{figure}

In Fig.~\ref{fig:heat_10ps_B2AR} we can find a snapshot of the protein
$\SI{10}{ps}$ after heating the ligand. In both states, energy reaches
TM3 and TM7 and travels in both directions along the backbone,
respectively. In the active state, more energy reaches TM7. We can
also see that some energy flows through the third contact in the
inactive state to TM5, but only a minor amount compared to the other
two helices. 

%However, we do not see notable correlation of energy transfer
%with structural changes\cite{Poudel21} or
%microswitches\cite{Fleetwood21} reported to be involved in protein
%activation.\cite{Rasmussen11} Differences in energy transfer rates are
%simply correlated with changes in contacts, which can be directly
%deduced from the involved crystal structure. 

Additionally, we calculated the energy transfer within a crystal structure
of an active $\beta_2$AR bound to adrenaline.\cite{Ring13} The results
are presented in Tabs.\ \SILigandContacts\ to \SIContactsActStates\
and Fig.\ \SIHeaterLigand. Although we observe differences in some
rates, especially a decrease of the transfer rate between adrenaline
and Asp113 by a factor of $\sim$2, the overall time course of energy
transfer is very similar to our results with the docked ligand.
As the hydrogen bond energy transfer forms the bottleneck step
in vibrational energy transport in $\beta_2$AR, the overall time course 
of the transfer is not affected as long as $D_B \gg D_C$.
This observation supporting our approach to evaluate energy transport within
$\beta_2$AR based on docking adrenaline into two receptor structures
that were not crystallized with this ligand. 

%\clearpage

Lastly, to check if a single crystal structure is sufficient to
predict the overall energy flow, we compared to energy transport
results obtained for a set of snapshots from MD simulations (see
Methods).  Tables~\SIContactsBetaARLong\ to
\SISecStructRatesBetaARLong\ show that the number of contacts
increases by a factor of $\sim$3, i.e., 141 (active) and 158 (inactive
receptor) contacts in MD simulations vs. 51 (active) and 66 (inactive
receptor) contacts in thee crystal structure. As the majority shows
only weak couplings, however, the additional contacts do not cause a
significant change of the time evolution of the energy distribution
(Fig.~\SIHeaterLigandBetaARLong). At least in the case of 7TM
proteins, using a single crystal structure seems sufficient to
reflect the overall vibrational energy transfer.

\vspace{-.6cm}
\section*{Discussion and Conclusion}
\vspace{-.6cm}

We have presented a computational approach that allows us to readily
compute the energy transport in any protein. As input we solely need a
protein structure, which e.g. can be deduced from crystal structural
data. Hence our master equation model may account for potential
structural heterogeneity only through an average over various
structures. We note that heterogeneous averaging should be appropriate,
because of the timescale separation between the fast (ps) energy
transport and the slow (ns to ms) conformational dynamics. Moreover,
the scaling rules rest on the assumption of a diffusive energy flow,
which is expected to be valid for a protein at room
temperature.\cite{Rubtsov19} Finally, we have focused on backbone and
contacts transport between standard residues, and have not
specifically parameterized specific cofactors, such as
chromophores. This should be sufficient, when we are interested in the
overall energy flow of the protein, which is the goal of this
paper. The model needs to be extended, when we want to compare to a
specific experiment as in Ref.\ \onlinecite{Deniz21}.

We applied the method to three different systems, bR, Rho and
$\beta_2$AR, all from the group of heptahelical transmembrane
proteins. The first two are photoreceptors, which are activated by the
absorption of an incoming photon, and are especially interesting to
consider as they need to dissipate the energy of the absorbed light
quantum. Even though the three systems appear similar in structure,
their energy transport mechanisms differ vastly.

bR manages to quickly direct excess energy out of the protein and thus
seems well suited to deal with the photon that it absorbs through the
retinal. As the two investigated states are very similar in their
secondary structure rates, they seem equally well suited for
dissipating excess energy out of the protein. The TM7 helix seems to
act like a ''lightning rod'' as it directs the energy out of the
protein and into the solvent. 

Rho, on the other hand, exhibits a rather poor energy dissipation:
excess energy mostly spreads along the backbone within TM7, but does
not quickly reach a part of the protein where it could dissipate into
a surrounding medium. Moreover, despite still covalently binding the
retinal cofactor, the active state seems less suited for quickly
distributing energy, which might potentially lead to overheating by
additionally absorbed photons. One may speculate that this problem is
potentially solved by the hydrolysis of the Schiff base between
protein and retinal and its subsequent dissociation.  

We finally considered $\beta_2$AR, which does not belong to the group
of photoreceptive proteins. When heating the ligand, the energy flows
into the protein via contacts with the ligand and is thus already much
more distributed than in the other systems, where only one entry way
into the main bulk of the protein is given. However, from these entry
points, the energy does not reach far and is consequently not
dissipated much in the protein. Because there is no photoexcitation
present in this protein, and adrenaline binding introduces a
negligible amount of energy into the protein, however, there is also
no need for an evolutionary adaption for the protein for optimal
energy dissipation.

Another reason to consider the energy flow in $\beta_2$AR is that its
energy transport pathways are believed to be linked to channels of
allosteric communication. \cite{Poudel21} Allostery may be mediated by
structural and dynamical changes of a
protein,\cite{Wodak19} and we indeed observed structural
differences between the inactive and active state of $\beta_2$AR.
These conformational changes are typically associated with changes of
interresidue contacts, which stabilize the two states. Since the
energy transport in a protein is also mediated by interresidue
contacts, various authors have studied potential interrelations of the
two processes. \cite{Ota05, Ishikura06, Nguyen09a, Martinez11,
  Ortiz15, Weber15, Maggi18} While it seems appealing to rationalize
elusive allosteric transitions via simple visualizations of vibrational
energy transport (as, e.g., shown in Figs.\ \ref{fig:heat_10ps_BR},
\ref{fig:heat_10ps_Rho} and \ref{fig:heat_10ps_B2AR}), there are
several limitations to this analogy.

First off, it should be stressed that energy transport and allostery
constitute different physical processes. Energy transport reflects the
instantaneous mechanical connectivity of the protein, and occurs
therefore on a picosecond timescale. Allostery, on the other hand, is
believed to reflect structural changes\cite{Nussinov15} (e.g., changes
of interresidue contacts or side chain dihedral angles) which
typically take place on a micro- or millisecond
timescale. \cite{Brueschweiler09,Buchli13, Bozovic20}
Secondly, our scaling rule for contact transport [Eq.\
(\ref{contact_rate})] provides a direct relation between protein
energy flow and polar contacts. Within this approximation, the contact
energy transport only gives information about the contacts that is
already known from the protein structure.
Finally, nonpolar contacts
are negligible for energy transport,\cite{Buchenberg16, Poudel20} but
are clearly important for allostery. In the case of $\beta_2$AR, for
example, Fleetwood et al.\cite{Fleetwood21} recently established the
importance of hydrophobic contacts forming microswitches.
While the energy flow may highlight aspects of the interresidue
couplings of a protein, it therefore seems not particularly suited to
explain allosteric phenomena.
\vspace*{-8mm}

\subsection*{Supplementary Information}
\vspace*{-5mm}
Ten Supplementary Tables detailing on energy transfer rates and three
Supplementary Figures displaying the energy distribution in protein
over time.  
\vspace*{-8mm}

%section*{Conflicts of interest}
%There are no conflicts to declare.

\subsection*{Acknowledgments}
\vspace*{-5mm} The authors thank Luis Valino-Borau for helpful
discussions. This work has been supported by the Deutsche
Forschungsgemeinschaft (DFG) via project STO247/10 and the Research
Unit FOR 5099 ''Reducing complexity of nonequilibrium'' (project
No. 431945604).
\vspace*{-2mm}

%\bibliography{\dir/stock,\dir/paper_new,\dir/md}
\bibliography{\dir/stock,\dir/md,new}

\end{document}

\author{Nadja Helmer}
	\author{Steffen Wolf}
	%\email[email:~]{steffen.wolf@physik.uni-freiburg.de}
	\author{Gerhard Stock}
	%\email[email:~]{stock@physik.uni-freiburg.de}
	\affiliation{Biomolecular Dynamics, Institute of Physics, Albert Ludwigs
		University, 79104 Freiburg, Germany.}
	\title{Supporting Information:\\Energy Transport and its Function in heptahelical Proteins}
	\date{\today}
	
	\maketitle

	\section{Supplementary Tables}

	\vspace{-.4cm}
	\begin{table}[h!]
	\caption{Contact between the different residues for the inactive and active state of bacteriorhodopsin. The rates are given in $\si{ps^{-1}}$. According to convention, the residues start at ID 5.}
	% residues start at 5 according to convention
	\label{tab:contacts_BR}
	\centering
	\begin{tabular}{llcc}
	\toprule
	\multicolumn{2}{c}{Residues} & \multicolumn{2}{c}{Rates}\\
	\midrule
	Res 1 &   Res 2 &  Inactive &   Active \\ 
	\cmidrule(lr){1-2}  \cmidrule(lr){3-4} 
	Arg7 &   Met60 &    0.13 &    0.11 \\
	Arg7 &   Leu61 &    0.14 &         \\
	Glu9 &   Tyr79 &    0.26 &    0.16 \\
	Leu13 &   Thr17 &    0.17 &    0.17 \\
	Met20 &   Thr24 &    0.15 &    0.18 \\
	Asp36 &  Gly231 &    0.09 &         \\
	Phe42 &   Thr46 &         &    0.15 \\
	Thr46 &   Asp96 &    0.16 &         \\
	Tyr57 &  Asp212 &    0.16 &    0.15 \\
	Ser59 &   Tyr64 &         &    0.26 \\
	Ser59 &   Gly65 &    0.10 &    0.09 \\
	Gly65 &   Ala81 &    0.14 &    0.10 \\
	Thr67 &   Ile78 &    0.40 &    0.35 \\
	Val69 &   Asn76 &    0.28 &    0.11 \\
	Phe71 &   Asn76 &    0.08 &         \\
	Asp85 &   Thr89 &         &    0.20 \\
	Trp86 &   Thr90 &    0.14 &    0.18 \\
	Trp86 &  Asp212 &         &    0.11 \\
	Ala98 &  Ala103 &    0.16 &    0.12 \\
	Asp102 &  Lys159 &    0.21 &         \\
	Ile117 &  Thr121 &    0.15 &    0.16 \\
	Met118 &  Ser141 &    0.14 &    0.13 \\
	Val124 &  Thr128 &    0.19 &    0.21 \\
	Ala126 &  Arg134 &    0.20 &    0.21 \\
	Thr128 &  Arg134 &    0.21 &    0.18 \\
	Arg134 &  Glu194 &    0.24 &    0.17 \\
	Trp138 &  Thr142 &    0.11 &    0.12 \\
	Trp138 &  Pro186 &    0.12 &    0.08 \\
	Arg164 &  Phe230 &    0.25 &         \\
	Pro165 &  Ser169 &    0.17 &    0.14 \\
	Glu166 &  Thr170 &    0.16 &         \\
	Leu174 &  Thr178 &         &    0.15 \\
	Val179 &  Ser183 &    0.14 &    0.15 \\
	Tyr185 &  Asp212 &    0.15 &    0.10 \\
	Leu190 &  Ala196 &    0.14 &    0.14 \\
	Ile191 &  Val199 &    0.14 &    0.15 \\
	Gly192 &  Gly197 &    0.15 &         \\
	Ser193 &  Glu204 &    0.44 &    0.35 \\
	Leu201 &  Thr205 &    0.15 &    0.09 \\
	Val210 &  Ser214 &         &    0.16 \\
	\bottomrule
	\end{tabular}
	\end{table}

	\begin{table*}[h!]
	\caption{Contact rates between the different residues for the inactive and active state of rhodopsin. The rates are given in $\si{ps^{-1}}$. The residue IDs start at 1.}
	% residues start at 1
	\label{tab:contacts_Rho}
	\centering
	\begin{minipage}{0.4\textwidth}
	\begin{tabular}{llcc}
	\toprule
	\multicolumn{2}{c}{Residues} & \multicolumn{2}{c}{Rates}\\
	\midrule
	Res 1 &   Res 2 &  Inactive &   Active \\ 
	\cmidrule(lr){1-2}  \cmidrule(lr){3-4} 
	Met1 &   Ser14 &    0.08 &         \\
	Asn2 &  Gly280 &    0.16 &         \\
	Thr4 &   Val11 &    0.42 &    0.16 \\
	Thr4 &   Phe13 &    0.25 &    0.26 \\
	Pro7 &  Arg177 &    0.21 &         \\
	Asn15 &   Val20 &    0.20 &         \\
	Arg21 &   Gln28 &    0.16 &    0.12 \\
	Ser22 &   Ala26 &    0.13 &    0.12 \\
	Pro23 &  Phe103 &    0.20 &         \\
	Gln28 &  Gly101 &         &    0.11 \\
	Tyr29 &  His100 &    0.11 &    0.10 \\
	Pro34 &   Ser38 &    0.11 &    0.11 \\
	Tyr43 &  Phe293 &    0.22 &         \\
	Gly51 &   Asn55 &    0.27 &    0.30 \\
	Ile54 &   Thr58 &    0.19 &    0.16 \\
	Asn55 &   Asp83 &    0.16 &    0.25 \\
	Asn55 &  Ala299 &    0.18 &    0.15 \\
	Thr58 &   Thr62 &    0.21 &    0.16 \\
	Tyr60 &   Gln64 &         &    0.08 \\
	Val61 &   His65 &         &    0.17 \\
	Thr62 &   Arg69 &    0.21 &         \\
	Val63 &   Arg69 &    0.19 &    0.23 \\
	Leu68 &   Asn73 &         &    0.19 \\
	Asn78 &  Trp161 &    0.22 &    0.13 \\
	Val87 &   Thr92 &         &    0.08 \\
	Phe88 &   Thr92 &    0.20 &         \\
	Gly89 &   Thr93 &         &    0.28 \\
	Gly90 &   Thr94 &         &    0.24 \\
	Thr93 &   Thr97 &    0.19 &         \\
	Thr94 &   Ser98 &    0.16 &    0.19 \\
	Tyr96 &  His100 &         &    0.15 \\
	Thr97 &  Tyr102 &         &    0.11 \\
	Thr97 &  Cys185 &         &    0.27 \\
	Glu113 &  Cys187 &    0.14 &    0.15 \\
	Glu113 &  Lys296 &    0.11 &         \\
	Gly114 &  Thr118 &    0.28 &    0.35 \\
	Glu122 &  His211 &    0.15 &         \\	
	\bottomrule
	\end{tabular}
	\end{minipage}
	\begin{minipage}{0.4\textwidth}
	\begin{tabular}{llcc}
	\toprule
	\multicolumn{2}{c}{Residues} & \multicolumn{2}{c}{Rates}\\
	\midrule
	Res 1 &   Res 2 &  Inactive &   Active \\ 
	\cmidrule(lr){1-2}  \cmidrule(lr){3-4}
	Ile123 &  Ser127 &         &    0.15 \\
	Arg135 &  Glu247 &    0.18 &         \\
	Met163 &  His211 &         &    0.13 \\
	Ala166 &  Tyr206 &    0.35 &    0.36 \\
	Ala168 &  Tyr178 &    0.31 &    0.35 \\
	Pro170 &  Trp175 &    0.15 &         \\
	Gly174 &  Thr198 &    0.22 &         \\
	Ser176 &  Thr198 &    0.24 &    0.19 \\
	Arg177 &  Asp190 &    0.23 &    0.36 \\
	Ile179 &  Gly188 &    0.21 &    0.17 \\
	Glu181 &  Ser186 &    0.15 &    0.20 \\
	Glu181 &  Tyr268 &    0.17 &    0.14 \\
	Thr193 &  Asn200 &    0.10 &    0.18 \\
	His195 &  Asn200 &    0.09 &    0.09 \\
	Gln225 &  Thr229 &    0.16 &         \\
	Lys231 &  Glu247 &         &    0.21 \\
	Ala234 &  Gln244 &         &    0.18 \\
	Ala235 &  Glu239 &    0.24 &         \\
	Ser240 &  Gln244 &         &    0.16 \\
	Glu247 &  Thr251 &    0.16 &    0.16 \\
	Glu249 &  Lys311 &    0.10 &         \\
	Phe273 &  Thr277 &         &    0.12 \\
	Tyr274 &  His278 &    0.13 &         \\
	Gly284 &  Met288 &         &    0.20 \\
	Pro285 &  Thr289 &    0.22 &    0.17 \\
	Phe293 &  Thr297 &         &    0.16 \\
	Phe294 &  Tyr301 &         &    0.17 \\
	Thr297 &  Tyr301 &    0.18 &         \\
	Ile307 &  Arg314 &    0.17 &         \\
	Met308 &  Arg314 &         &    0.15 \\
	Met309 &  Arg314 &         &    0.10 \\
	Lys311 &  Asn315 &    0.18 &         \\
	Gln312 &  Thr335 &    0.10 &         \\
	Asn315 &  Thr319 &         &    0.13 \\
	Cys316 &  Thr320 &    0.28 &    0.16 \\
	Ser338 &  Ser343 &    0.36 &         \\
	\bottomrule
	\end{tabular}
	\end{minipage}
	\end{table*}

\begin{table}[h!]
	\centering
	\caption{Rates of Rhodopsin with surrounding water molecules. In the inactive state, we find 37 contacts with water (yielding 28 contact rates). In the active state we only identify 17 water contacts (16 contact rates). The rates are given in \si{ps^{-1}}.}
	\label{rho_contacts_water}
	\begin{minipage}[t]{0.3\textwidth}
		\begin{tabular}[t]{llrr}
			\toprule
			\multicolumn{2}{c}{Residues} & \multicolumn{2}{c}{Rates}\\ \cmidrule(lr){1-2}  \cmidrule(lr){3-4}  
			Res 1 &  Res 2 &  Inactive  &   Active    \\	\midrule
			Gly3 &    Water &    0.16   &    0.16  \\
			Glu5 &    Water &           &    0.09  \\
			Asn8 &    Water &    0.08   &          \\
			Tyr10 &    Water &    0.11  &          \\
			Pro12 &    Water &    0.08  &          \\
			Ser14 &    Water &    0.09  &    0.15  \\
			Pro34 &    Water &          &    0.11  \\
			Gln64 &    Water &          &    0.06  \\
			Asn73 &    Water &    0.10  &          \\
			Gly89 &    Water &    0.20  &          \\
			Ser98 &    Water &          &    0.11  \\
			Glu113 &    Water &    0.08 &          \\
			Gly120 &    Water &    0.07 &          \\
			Ala153 &    Water &         &    0.12  \\
			Pro171 &    Water &    0.11 &          \\
			Ser176 &    Water &    0.11 &          \\
			Tyr178 &    Water &    0.15 &          \\
			Pro180 &    Water &    0.14 &          \\
			Glu181 &    Water &    0.15 &    0.15  \\
			\bottomrule
		\end{tabular}
	\end{minipage}
	\begin{minipage}[t]{0.3\textwidth}
		\begin{tabular}[t]{llrr}
			\toprule
			\multicolumn{2}{c}{Residues} & \multicolumn{2}{c}{Rates}\\ \cmidrule(lr){1-2}  \cmidrule(lr){3-4}  
			Res 1 &  Res 2 &  Inactive  &   Active    \\	\midrule
			
			Cys185 &    Water &         &    0.06  \\
			Cys187 &    Water &    0.07 &          \\
			Tyr191 &    Water &    0.04 &          \\
			Tyr192 &    Water &    0.08 &          \\
			Thr193 &    Water &    0.07 &          \\
			Glu196 &    Water &    0.10 &          \\
			Thr229 &    Water &         &    0.10  \\
			Gln236 &    Water &         &    0.07  \\
			Glu239 &    Water &         &    0.09  \\
			Glu249 &    Water &    0.05 &          \\
			Met257 &    Water &    0.09 &    0.12  \\
			Ala260 &    Water &         &    0.10  \\
			Cys264 &    Water &    0.12 &          \\
			Gly284 &    Water &    0.27 &          \\
			Pro291 &    Water &    0.08 &          \\
			Ser298 &    Water &    0.23 &    0.07  \\
			Asn302 &    Water &    0.08 &          \\
			Cys323 &    Water &         &    0.05  \\
			Asp331 &    Water &    0.04 &          \\
			Glu332 &    Water &    0.11 &          \\
			
			\bottomrule
		\end{tabular}
	\end{minipage}
\end{table}

	\begin{table*}[h!]
	\caption{Contact rates for the inactive and active state of $\beta_2$AR. The rates are given in $\si{ps^{-1}}$. According to convention, the residue IDs start at 29.}
	% residues start at 29 according to convention
	\label{tab:contacts_B2AR}
	\centering
	\begin{minipage}{0.4\textwidth}
	\begin{tabular}{llcc}
	\toprule
	\multicolumn{2}{c}{Residues} & \multicolumn{2}{c}{Rates}\\
	\midrule
	Res 1 &   Res 2 &  Inactive &   Active \\ 
	\cmidrule(lr){1-2}  \cmidrule(lr){3-4} 
	Glu30 &   Lys97 &         &    0.25 \\
	Gly37 &   Ser41 &    0.24 &    0.27 \\
	Ile47 &   Asn51 &    0.13 &    0.18 \\
	Asn51 &  Ser319 &    0.11 &    0.17 \\
	Val52 &   Thr56 &    0.21 &    0.24 \\
	Ala57 &   Phe61 &         &    0.25 \\
	Ala59 &   Gln65 &    0.21 &         \\
	Leu64 &   Asn69 &    0.17 &    0.16 \\
	Thr68 &  Asp130 &    0.21 &         \\
	Asn69 &   Thr73 &    0.11 &    0.15 \\
	Asp79 &  Ser120 &         &    0.16 \\
	Asp79 &  Ser319 &    0.12 &         \\
	Val86 &  Trp109 &    0.19 &    0.24 \\
	His93 &  Cys191 &    0.15 &         \\
	Trp99 &  Thr189 &    0.11 &         \\
	Asn103 &  Ala186 &         &    0.11 \\
	Cys106 &  Thr110 &    0.23 &    0.24 \\
	Glu107 &  Ser111 &    0.18 &    0.19 \\
	Glu107 &  Gln170 &         &    0.19 \\
	Glu107 &  Tyr174 &    0.30 &         \\
	Glu107 &  Arg175 &    0.41 &         \\
	Asp113 &  Tyr316 &    0.18 &    0.28 \\
	Asp113 &  Adr343 &    0.71 &    0.57 \\
	Val114 &  Thr118 &    0.19 &    0.17 \\
	Cys116 &  Ser120 &    0.23 &         \\
	Ala119 &  Thr123 &    0.26 &    0.17 \\
	Glu122 &  Val206 &         &    0.17 \\
	Asp130 &  Ser143 &    0.21 &         \\
	Arg131 &  Tyr141 &    0.15 &         \\
	Arg131 &  Gln142 &    0.16 &         \\
	Tyr132 &  Thr136 &    0.12 &    0.13 \\
	Ala134 &  Tyr141 &    0.17 &         \\
	Lys140 &  Gln243 &    0.18 &         \\
	Lys147 &  Arg151 &    0.17 &         \\
	Val157 &  Ser161 &    0.16 &    0.20 \\
	Val160 &  Thr164 &         &    0.20 \\
	Ser161 &  Ser165 &    0.20 &    0.21 \\
	Thr164 &  Tyr199 &         &    0.14 \\
	Pro168 &  Tyr174 &    0.20 &         \\
	Trp173 &  Asn196 &         &    0.16 \\
	Tyr174 &  Asn196 &    0.10 &         \\
	Arg175 &  Asn196 &    0.14 &    0.12 \\
	Ala176 &  Tyr185 &    0.34 &         \\

	\bottomrule
	\end{tabular}
	\end{minipage}
	\begin{minipage}{0.4\textwidth}
	\begin{tabular}{llcc}
	\toprule
	\multicolumn{2}{c}{Residues} & \multicolumn{2}{c}{Rates}\\
	\midrule
	Res 1 &   Res 2 &  Inactive &   Active \\ 
	\cmidrule(lr){1-2}  \cmidrule(lr){3-4}
	Ala176 &  Phe194 &         &    0.18 \\
	Gln179 &  Asn183 &         &    0.14 \\
	Tyr185 &  Phe194 &         &    0.13 \\
	Asp192 &  Lys305 &    0.24 &    0.26 \\
	Tyr199 &  Ser203 &         &    0.15 \\
	Ala200 &  Ser204 &    0.17 &    0.18 \\
	Ser203 &  Adr343 &    0.29 &         \\
	Tyr209 &  Phe290 &    0.14 &    0.19 \\
	Val216 &  Ser220 &         &    0.21 \\
	Tyr219 &  Leu272 &    0.15 &         \\
	Arg221 &  Glu225 &         &    0.18 \\
	Glu225 &  Gln229 &         &    0.18 \\
	Arg228 &  Glu249 &    0.34 &         \\
	Gln231 &  Glu268 &         &    0.09 \\
	Asp234 &  Arg239 &    0.52 &         \\
	Asp234 &  Arg260 &    0.55 &         \\
	Ser236 &  Asp251 &         &    0.48 \\
	Gly238 &  Asn244 &         &    0.21 \\
	Arg239 &  His256 &    0.13 &         \\
	Phe240 &  Thr254 &    0.18 &         \\
	His241 &  Gly252 &    0.10 &         \\
	Val242 &  Gln250 &    0.17 &         \\
	Gln243 &  Glu249 &    0.13 &         \\
	Asn244 &  Val248 &    0.18 &         \\
	Glu249 &  Arg253 &         &    0.43 \\
	Lys270 &  Thr274 &         &    0.15 \\
	Lys270 &  Arg328 &    0.13 &         \\
	Ile277 &  Thr281 &    0.20 &    0.15 \\
	Ile278 &  Tyr326 &         &    0.24 \\
	Met279 &  Thr283 &    0.22 &    0.20 \\
	Trp286 &  Asn318 &         &    0.09 \\
	Phe289 &  Asn293 &    0.16 &         \\
	Asn293 &  Tyr308 &    0.19 &    0.22 \\
	Val295 &  Gln299 &         &    0.14 \\
	His296 &  Asn301 &    0.12 &         \\
	His296 &  Ile303 &    0.09 &         \\
	Asn312 &  Adr343 &    0.27 &    0.27 \\
	Gly315 &  Ser319 &    0.11 &         \\
	Tyr316 &  Gly320 &    0.08 &         \\
	Asn318 &  Asn322 &    0.15 &    0.22 \\
	Cys327 &  Arg333 &    0.23 &         \\
	Arg328 &  Arg333 &         &    0.20 \\
	Gln337 &  Leu342 &    0.17 &         \\
	\bottomrule
	\end{tabular}
	\end{minipage}
	\end{table*}

\begin{table}[h!]
	\caption{Rates of the contacts formed between the ligand adrenaline and $\beta_2$AR for the 4LDO state and the 3P0G state. The rates are given in $\si{ps^{-1}}$.}
	\label{tab:B2AR_ligand_contacts}
	\centering
	\begin{tabular}[t]{cccc}
		\toprule
		\multicolumn{2}{c}{Residues} & \multicolumn{2}{c}{Rates}\\ 
		\midrule
		Res 1 & Res 2 & 4LDO & 3P0G \\ \cmidrule(lr){1-2}  \cmidrule(lr){3-4}  
		Adr  & Asp113 &    0.11 &     0.27 \\	
		Adr  & Asn312 &    0.13 &      0.14 \\
		\bottomrule
	\end{tabular}
\end{table}

\begin{table}[h!]
	\caption{Secondary structure rates for both active states, 4LDO (left) and 3P0G (right). The rates are given in $\si{ps^{-1}}$.}
	\label{tab:sec_rates}
	\small
	\centering
	\begin{tabular}{llrr}
		\toprule
		\multicolumn{2}{c}{Sec. Struct.} & \multicolumn{2}{c}{Rates}\\ 
		\midrule
		Sec 1 &    Sec 2 &  4LDO &  3P0G \\
		\midrule
		N-Term &  ECL1   &         &    0.25 \\
		TM1    &  TM7    &    0.11 &    0.17 \\
		ICL1   &  H8     &    0.09 &         \\
		TM2    &  TM3    &    0.49 &    0.40 \\
		TM3    &  TM4    &         &    0.19 \\
		TM3    &  ECL2   &    1.03 &    0.11 \\
		TM3    &  TM5    &    0.19 &    0.17 \\
		TM3    &  TM7    &    0.24 &    0.28 \\
		TM4    &  TM5    &    0.21 &    0.14 \\
		ECL2   &  TM7    &         &    0.26 \\
		TM5    &  TM6    &    0.12 &    0.19 \\
		TM6    &  TM7    &    0.29 &    0.54 \\
		\bottomrule
	\end{tabular}
\end{table}

\begin{table}[h!]
	\caption{Contact rates for 4LDO active state of $\beta_2$AR. The rates are given in $\si{ps^{-1}}$. According to convention, the residue IDs start at 29.}
	% residues start at 29 according to convention
	\label{tab:contacts_B2AR_act}
	\centering
	\begin{minipage}[t]{0.4\textwidth}
		\begin{tabular}[t]{llcc}
			\toprule
			\multicolumn{2}{c}{Residues} & \multicolumn{2}{c}{Rates}\\
			\midrule
			Res 1 &   Res 2 & 4LDO & 3P0G\\ 
			\cmidrule(lr){1-2}  \cmidrule(lr){3-4} 
			Glu30 &   Lys97 &         &    0.25 \\
			Gly37 &   Ser41 &    0.16 &    0.27 \\
			Ile47 &   Asn51 &    0.15 &    0.18 \\
			Asn51 &  Ser319 &    0.11 &    0.17 \\
			Val52 &   Thr56 &    0.15 &    0.24 \\
			Ala57 &   Phe61 &         &    0.25 \\
			Arg63 &  Asp331 &    0.09 &         \\
			Leu64 &   Asn69 &    0.15 &    0.16 \\
			Thr68 &  Asp130 &    0.12 &         \\
			Asn69 &   Thr73 &    0.15 &    0.15 \\
			Asp79 &  Ser120 &    0.14 &    0.16 \\
			Val86 &  Trp109 &    0.23 &    0.24 \\
			Ala92 &   Met96 &    0.23 &         \\
			Asn103 &  Ala186 &         &    0.11 \\
			Asn103 &  Gln187 &    0.16 &         \\
			Asn103 &  Glu188 &    0.22 &         \\
			Cys106 &  Thr110 &    0.17 &    0.24 \\
			Glu107 &  Ser111 &    0.12 &    0.19 \\
			Glu107 &  Gln170 &         &    0.19 \\
			Glu107 &  Tyr174 &    0.23 &         \\
			Glu107 &  Arg175 &    0.42 &         \\
			Asp113 &  Tyr316 &    0.24 &    0.28 \\
			Asp113 &  Adr343 &    0.22 &    0.57 \\
			Val114 &  Thr118 &    0.18 &    0.17 \\
			Ala119 &  Thr123 &    0.19 &    0.17 \\
			Glu122 &  Val206 &    0.09 &    0.17 \\
			Asp130 &  Tyr141 &    0.22 &         \\
			Tyr132 &  Thr136 &         &    0.13 \\
			Tyr132 &  Glu225 &    0.09 &         \\
			Leu144 &  Lys149 &    0.22 &         \\
			Val157 &  Ser161 &    0.13 &    0.20 \\
			Val160 &  Thr164 &         &    0.20 \\
			Ser161 &  Ser165 &    0.18 &    0.21 \\
			Thr164 &  Tyr199 &    0.21 &    0.14 \\
			Pro168 &  Trp173 &    0.24 &         \\
			Pro168 &  Tyr174 &    0.14 &         \\
			Trp173 &  Asn196 &    0.09 &    0.16 \\
			
			\bottomrule
		\end{tabular}
	\end{minipage}
	\begin{minipage}[t]{0.3\textwidth}
		\begin{tabular}[t]{llcc}
			\toprule
			\multicolumn{2}{c}{Residues} & \multicolumn{2}{c}{Rates}\\
			\midrule
			Res 1 &   Res 2 & 4LDO & 3P0G\\ 
			\cmidrule(lr){1-2}  \cmidrule(lr){3-4} 
			
			Tyr174 &  Asn196 &    0.09 &         \\
			Arg175 &  Asn196 &    0.14 &    0.12 \\
			Ala176 &  Phe194 &         &    0.18 \\
			Gln179 &  Asn183 &    0.17 &    0.14 \\
			Tyr185 &  Phe194 &         &    0.13 \\
			Asp192 &  Lys305 &         &    0.26 \\
			Tyr199 &  Ser203 &         &    0.15 \\
			Ala200 &  Ser204 &         &    0.18 \\
			Ser203 &  Ser207 &    0.19 &         \\
			Tyr209 &  Phe290 &    0.12 &    0.19 \\
			Val216 &  Ser220 &         &    0.21 \\
			Arg221 &  Glu225 &    0.07 &    0.18 \\
			Glu225 &  Gln229 &    0.22 &    0.18 \\
			Gln231 &  Lys235 &    0.19 &         \\
			Gln231 &  Glu268 &         &    0.09 \\
			Asp234 &  Arg239 &    0.31 &         \\
			Lys235 &  Ser262 &    0.08 &         \\
			Ser236 &  Asp251 &         &    0.48 \\
			Glu237 &  Arg253 &    0.20 &         \\
			Gly238 &  Asn244 &         &    0.21 \\
			Glu249 &  Arg253 &         &    0.43 \\
			Gly255 &  Ser262 &    0.17 &         \\
			Leu258 &  Ser262 &    0.07 &         \\
			Lys270 &  Thr274 &    0.12 &    0.15 \\
			Ile277 &  Thr281 &    0.11 &    0.15 \\
			Ile278 &  Tyr326 &         &    0.24 \\
			Met279 &  Thr283 &         &    0.20 \\
			Trp286 &  Asn318 &         &    0.09 \\
			Asn293 &  Tyr308 &    0.29 &    0.22 \\
			Val295 &  Gln299 &         &    0.14 \\
			His296 &  Asn301 &    0.08 &         \\
			Asn312 &  Adr343 &    0.24 &    0.27 \\
			Asn318 &  Asn322 &         &    0.22 \\
			Asn322 &  Tyr326 &    0.19 &         \\
			Arg328 &  Arg333 &         &    0.20 \\
			Gln337 &  Leu342 &    0.14 &         \\
			\bottomrule
		\end{tabular}
	\end{minipage}
\end{table}

\newpage

	\begin{table*}[h!]
	\caption{Contact rates between the different residues for the inactive and active state of $\beta_2$AR based on 7 snapshots from the trajectory. The rates are given in $\si{ps^{-1}}$. The residue IDs start at 1.}
	% residues start at 1
	\label{tab:contacts_B2AR_long}
	\centering
	\begin{minipage}{0.3\textwidth}
		\begin{tabular}{llcc}
			\toprule
			\multicolumn{2}{c}{Residues} & \multicolumn{2}{c}{Rates}\\
			\midrule
			Res 1 &   Res 2 &  Inactive &   Active \\ 
			\cmidrule(lr){1-2}  \cmidrule(lr){3-4} 
            Asp29 &  Val33 &          &    0.02 \\
            Glu29 &  Lys97 &          &    0.04 \\
            Gly37 &  Ser41 &    0.22 &     0.17  \\
            Ile47 &  Asn51 &    0.15 &     0.16  \\
            Asn51 & Ser319 &    0.13 &     0.12  \\
            Val52 &  Thr56 &    0.21 &     0.19  \\
            Ala57 &  Phe61 &         &     0.21  \\
            Ala57 &  Gln65 &    0.03  &         \\
            Ile58 &  Gln65 &    0.15 &     0.11  \\
            Arg63 & Asp331 &         &     0.15  \\
            Arg63 & Glu338 &    0.14 &           \\
            Leu64 &  Asn69 &    0.13 &     0.08  \\
            Gln65 &  Tyr70 &          &    0.02 \\
            Thr68 & Asp130 &    0.19 &     0.03  \\
            Asn69 &  Thr73 &    0.18 &     0.18  \\
            Asn69 & Tyr326 &    0.08 &           \\
            Tyr70 &  Ser74 &    0.11 &           \\
            Tyr70 & Ala150 &    0.07 &           \\
            Asp79 & Ser120 &         &     0.06  \\
            Asp79 & Ser319 &    0.06 &           \\
            Asp79 & Asn322 &          &    0.05 \\
            Val86 & Trp109 &    0.21 &     0.22  \\
            Gly90 & Trp313 &    0.09 &           \\
            His93 &  Met98 &    0.01  &         \\
            His93 & Cys191 &    0.12 &     0.07  \\
            Lys97 & Glu306 &    0.05  &         \\
            Trp99 & Thr189 &    0.05 &     0.07  \\
            Asn103 & Glu107 &         &    0.02 \\
            Asn103 & Tyr185 &    0.12 &          \\
            Asn103 & Ala186 &         &    0.02 \\
            Asn103 & Gln187 &    0.06 &          \\
            Asn103 & Glu188 &    0.06 &          \\
            Cys106 & Thr110 &    0.24 &     0.24 \\
            Glu107 & Ser111 &    0.18 &     0.18 \\
            Glu107 & Gln170 &         &    0.03 \\
            Glu107 & His172 &    0.07 &     0.15 \\
            Glu107 & Tyr174 &    0.27 &          \\
            Glu107 & Arg175 &    0.35 &     0.07 \\
            Asp113 & Asn312 &    0.01 &         \\
            Asp113 & Tyr316 &    0.27 &     0.26 \\
            Val114 & Thr118 &    0.17 &     0.19 \\
            Cys116 & Ser120 &    0.20 &     0.07 \\
            Ala119 & Thr123 &    0.24 &     0.24 \\
            Glu122 & Ser161 &    0.02 &         \\
            Glu122 & Val206 &         &     0.20 \\
            Asp130 & Tyr141 &         &    0.04 \\
            Asp130 & Ser143 &    0.20 &          \\
            Arg131 & Tyr141 &    0.02 &         \\
            Arg131 & Gln142 &    0.14 &          \\
            Tyr132 & Thr136 &    0.06 &     0.16 \\
            Phe133 & Ser137 &    0.14 &          \\
            Ala134 & Tyr141 &    0.02 &         \\
            Ile135 & Gln229 &    0.12 &          \\

			\bottomrule
		\end{tabular}
	\end{minipage}
	\begin{minipage}{0.3\textwidth}
		\begin{tabular}{llcc}
			\toprule
			\multicolumn{2}{c}{Residues} & \multicolumn{2}{c}{Rates}\\
			\midrule
			Res 1 &   Res 2 &  Inactive &   Active \\ 
			\cmidrule(lr){1-2}  \cmidrule(lr){3-4}
            
            Ile135 & Gln243 &    0.01 &         \\
            Thr136 & Gln229 &         &    0.02 \\
            Pro138 & Gln142 &         &    0.05 \\
            Lys140 & Gln243 &    0.05 &         \\
            Tyr141 & Glu268 &    0.10 &          \\
            Ser143 & Lys149 &         &    0.04 \\
            Lys147 & Arg151 &    0.05 &         \\
            Val157 & Ser161 &    0.10 &     0.14 \\
            Val160 & Thr164 &    0.16 &     0.19 \\
            Ser161 & Ser165 &    0.14 &     0.22 \\
            Thr164 & Tyr199 &         &     0.18 \\
            Pro168 & Trp173 &    0.03 &         \\
            Pro168 & Tyr174 &    0.09 &     0.07 \\
            Trp173 & Asn196 &    0.08 &     0.10 \\
            Tyr174 & Asn196 &    0.11 &     0.10 \\
            Arg175 & Asn196 &    0.07 &     0.09 \\
            Ala176 & Tyr185 &    0.05 &         \\
            Ala176 & Phe194 &    0.12 &     0.20 \\
            His178 & Val297 &         &    0.02 \\
            Gln179 & Asn183 &    0.03 &    0.02 \\
            Asn183 & Gln187 &    0.05 &         \\
            Tyr185 & Phe194 &    0.17 &     0.14 \\
            Asp192 & Lys305 &    0.17 &     0.22 \\
            Phe193 & Lys305 &    0.03 &         \\
            Gln197 & Ile294 &         &    0.04 \\
            Tyr199 & Ser203 &    0.08 &     0.15 \\
            Ala200 & Ser204 &    0.21 &     0.20 \\
            Ser203 & Ser207 &    0.12 &          \\
           
            Tyr209 & Phe290 &    0.17 &     0.17 \\
            Val216 & Ser220 &    0.15 &     0.17 \\
            Tyr219 & Leu272 &    0.15 &          \\
            Arg221 & Glu225 &         &     0.14 \\
            Glu225 & Gln229 &    0.15 &     0.28 \\
            Lys227 & Gln231 &    0.04 &         \\
            Lys227 & Glu268 &         &    0.13  \\
            Arg228 & Glu249 &    0.25 &          \\
            Gln229 & Ser262 &    0.09 &          \\
            Gln229 & Lys263 &    0.06 &          \\
            Leu230 & Arg239 &    0.16 &          \\
            Gln231 & Arg239 &    0.03 &         \\
            Gln231 & Phe240 &    0.02 &         \\
            Gln231 & His241 &    0.02 &         \\ 
            Gln231 & Glu268 &         &    0.04 \\ 
            Lys232 & Gln250 &         &    0.02 \\
            Lys232 & Ser262 &         &     0.11 \\
            Ile233 & Gly238 &    0.06 &          \\
            Asp234 & Arg239 &    0.18 &          \\
            Asp234 & Gln250 &         &    0.02 \\
            Asp234 & Arg253 &         &     0.30 \\
            Asp234 & Thr254 &         &    0.03 \\
            Asp234 & Arg260 &    0.08 &     0.11 \\
            Lys235 & Arg260 &         &    0.03 \\
           Ser236 & Gln250 &         &     0.14 \\
        \bottomrule            
	\end{tabular}
\end{minipage}
\begin{minipage}{0.3\textwidth}
\begin{tabular}{llcc}
           \toprule
           \multicolumn{2}{c}{Residues} & \multicolumn{2}{c}{Rates}\\
           \midrule
           Res 1 &   Res 2 &  Inactive &   Active \\ 
              \cmidrule(lr){1-2}  \cmidrule(lr){3-4}
    
            Ser236 & Asp251 &         &     0.07 \\
            Ser236 & Arg253 &    0.07 &          \\
            Ser236 & Arg260 &         &    0.04 \\
            Glu237 & Arg253 &    0.10 &     0.05 \\
            Glu237 & Arg260 &    0.26 &     0.09 \\
            Gly238 & Asn244 &         &    0.03 \\
            Arg239 & Asn244 &         &    0.02 \\
            Arg239 & His256 &    0.02 &         \\
            Arg239 & Arg259 &    0.12 &          \\
            Arg239 & Arg260 &    0.07 &          \\
            Phe240 & Thr254 &    0.03 &         \\
            His241 & Gly252 &    0.06 &          \\
            Val242 & Gln250 &    0.08 &          \\
            Val242 & Gly252 &    0.01 &         \\
            Gln243 & Glu249 &    0.07 &          \\
            Asn244 & Val248 &    0.12 &          \\
            Gln247 & Arg253 &         &    0.03 \\
            Val248 & Arg253 &         &    0.02 \\
            Glu249 & Arg253 &         &     0.21 \\
            Gln250 & Arg260 &         &     0.08 \\
            Gln250 & Thr254 &    0.05 &    0.10 \\
            Gly255 & Arg259 &         &    0.03 \\
            Arg253 & Arg260 &         &    0.03 \\
            Gly255 & Arg260 &         &    0.04 \\
            Gly257 & Leu266 &         &    0.03 \\
            Arg259 & Lys263 &         &    0.02 \\
            Arg259 & Phe264 &         &    0.03 \\
            Arg259 & Cys265 &    0.01 &         \\
            Cys265 & His269 &    0.08 &          \\
            Lys267 & Asp331 &    0.09 &          \\
            Lys270 & Thr274 &         &    0.05 \\
            Lys270 & Arg328 &    0.02 &         \\
            Lys270 & Asp331 &    0.03 &         \\
            Ile277 & Thr281 &    0.14 &     0.15 \\
            Ile278 & Tyr326 &         &    0.04 \\
            Met279 & Thr283 &    0.18 &     0.15 \\
            Thr281 & Asn318 &         &     0.09 \\
            Trp286 & Asn318 &         &     0.06 \\
            Phe289 & Asn293 &    0.02 &         \\
            Phe289 & Tyr308 &    0.03 &         \\
            Asn293 & Tyr308 &    0.03 &     0.24 \\
            Val295 & Gln299 &         &    0.02 \\
            His296 & Asn301 &    0.02 &         \\
            His296 & Ile303 &    0.01 &         \\
            Gly315 & Ser319 &    0.19 &          \\
            Tyr316 & Gly320 &    0.03 &         \\
            Asn318 & Asn322 &    0.15 &     0.16 \\
            Leu324 & Arg328 &         &     0.16 \\
            Cys327 & Arg333 &    0.18 &          \\
            Arg328 & Arg333 &         &     0.20 \\
            Ser329 & Arg333 &         &    0.03 \\
            Arg333 & Gln337 &    0.09 &          \\
            Gln337 & Leu342 &    0.16 &     0.02 \\
			\bottomrule
		\end{tabular}
	\end{minipage}
\end{table*}

\begin{table}[h!]
	\caption{Rates of the contacts formed between the ligand adrenaline and $\beta_2$AR for the active and the inactive state using 7 frames from the trajectory. The rates are given in $\si{ps^{-1}}$.}
	\label{tab:ligand_contacts_B2AR_long}
	\centering
	\begin{tabular}[t]{cccc}
		\toprule
		\multicolumn{2}{c}{Residues} & \multicolumn{2}{c}{Rates}\\ 
		\midrule
		Res 1 & Res 2 & Inactive & Active \\ \cmidrule(lr){1-2}  \cmidrule(lr){3-4}  
		Adr & Asp113 &    0.29 &     0.24 \\
        Adr & Ser203 &    0.02 &         \\
        Adr & Asn312 &    0.14 &     0.15 \\
		\bottomrule
	\end{tabular}
\end{table}

\begin{table}[h!]
	\caption{Secondary structure rates for the active and the inactive state of $\beta_2$AR based on 7 snapshot from the simulation trajectory. The rates are given in $\si{ps^{-1}}$.}
	\label{tab:sec_rates_B2AR_long}
	\small
	\centering
	\begin{tabular}{llrr}
		\toprule
		\multicolumn{2}{c}{Sec. Struct.} & \multicolumn{2}{c}{Rates}\\ 
		\midrule
		Sec 1 &    Sec 2 &  Inactive &  Active \\
		\midrule
		N-Term &  ECL1   &         &    0.04 \\
		TM1    &  TM7    &    0.13 &    0.12 \\
		ICL1   &  H8     &    0.14 &    0.15 \\
		TM2    &  TM3    &    0.40 &    0.31 \\
		TM2    &  TM4    &    0.07 &         \\
		TM2    &  ECL2   &    0.12 &    0.06 \\
		TM2    &  TM7    &    0.23 &    0.05 \\
		ECL1   &  ECL2   &    0.05 &    0.07 \\
		ECL1   &  TM7    &    0.05 &         \\
		TM3    &  TM4    &    0.02 &    0.03 \\
		TM3    &  ECL2   &    0.92 &    0.24 \\
		TM3    &  TM5    &    0.12 &    0.20 \\
		TM3    &  ICL3   &    0.01 &    0.02 \\
		TM3    &  TM7    &    0.28 &    0.26 \\
		ICL2   &  ICL3   &    0.05 &         \\
		ICL2   &  TM6    &    0.10 &         \\
		TM4    &  TM5    &         &    0.18 \\
		ECL2   &  TM6    &         &    0.02 \\
		ECL2   &  TM7    &    0.20 &    0.22 \\
		TM5    &  TM6    &    0.32 &    0.34 \\
		
		TM6    &  TM7    &    0.06 &    0.44 \\
		TM6    &  ICL4   &    0.02 &         \\
		TM6    &  H8     &    0.11 &         \\
		TM7    &  H8     &    0.18 &         \\
	
		\bottomrule
	\end{tabular}
\end{table}

\newpage
\section{Supplementary Figures}

\begin{figure}[h]
	\centering
	\includegraphics[width=.6\linewidth]{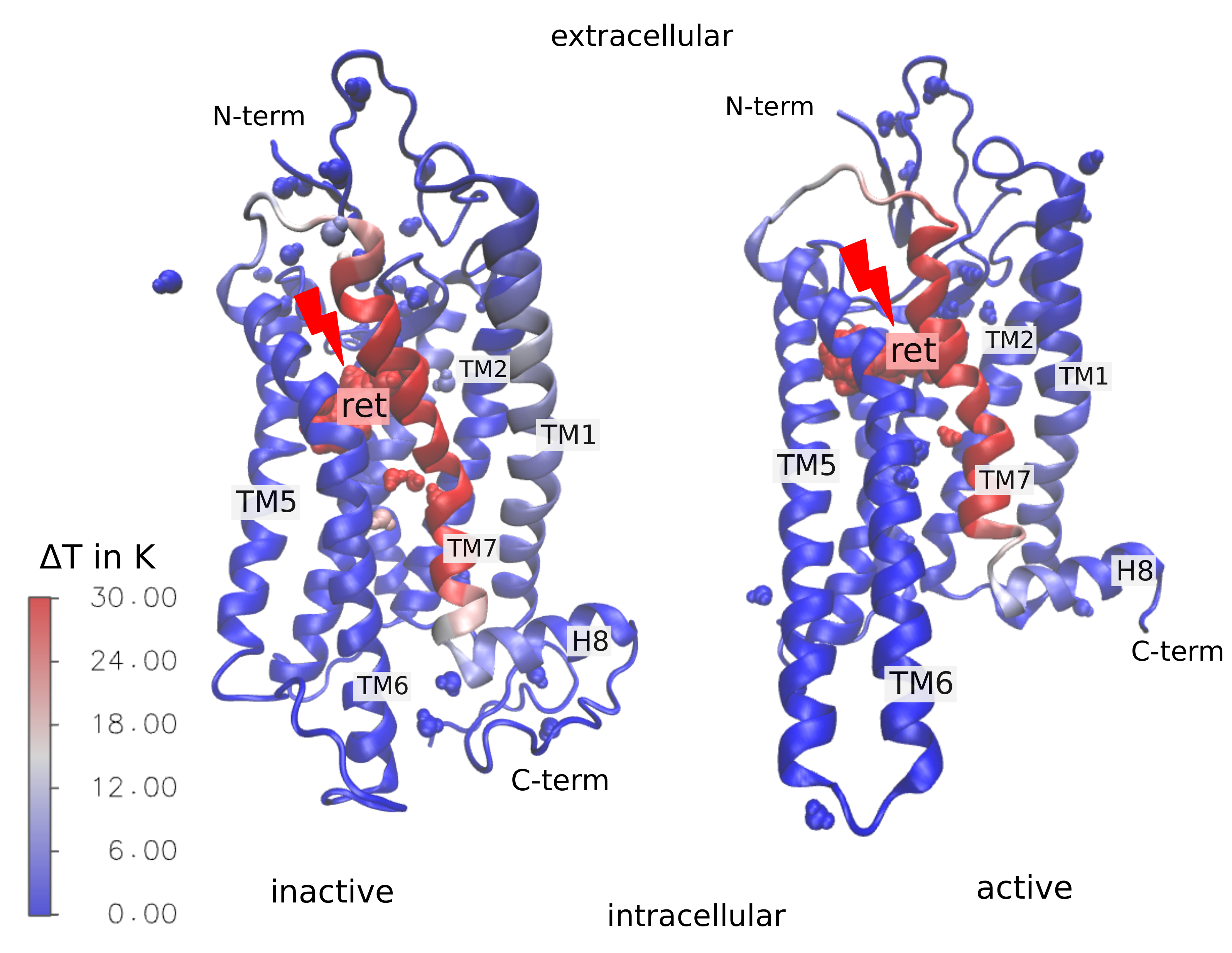}
	\caption{Rhodopsin shown in the inactive and active state including water molecules \SI{10}{ps} after heating the retinal. The water molecules are depicted as spheres.}
	\label{fig:Rho_heaterRet_withwater}
\end{figure}

\begin{figure}[h]
	\centering
	\includegraphics[width=.6\linewidth]{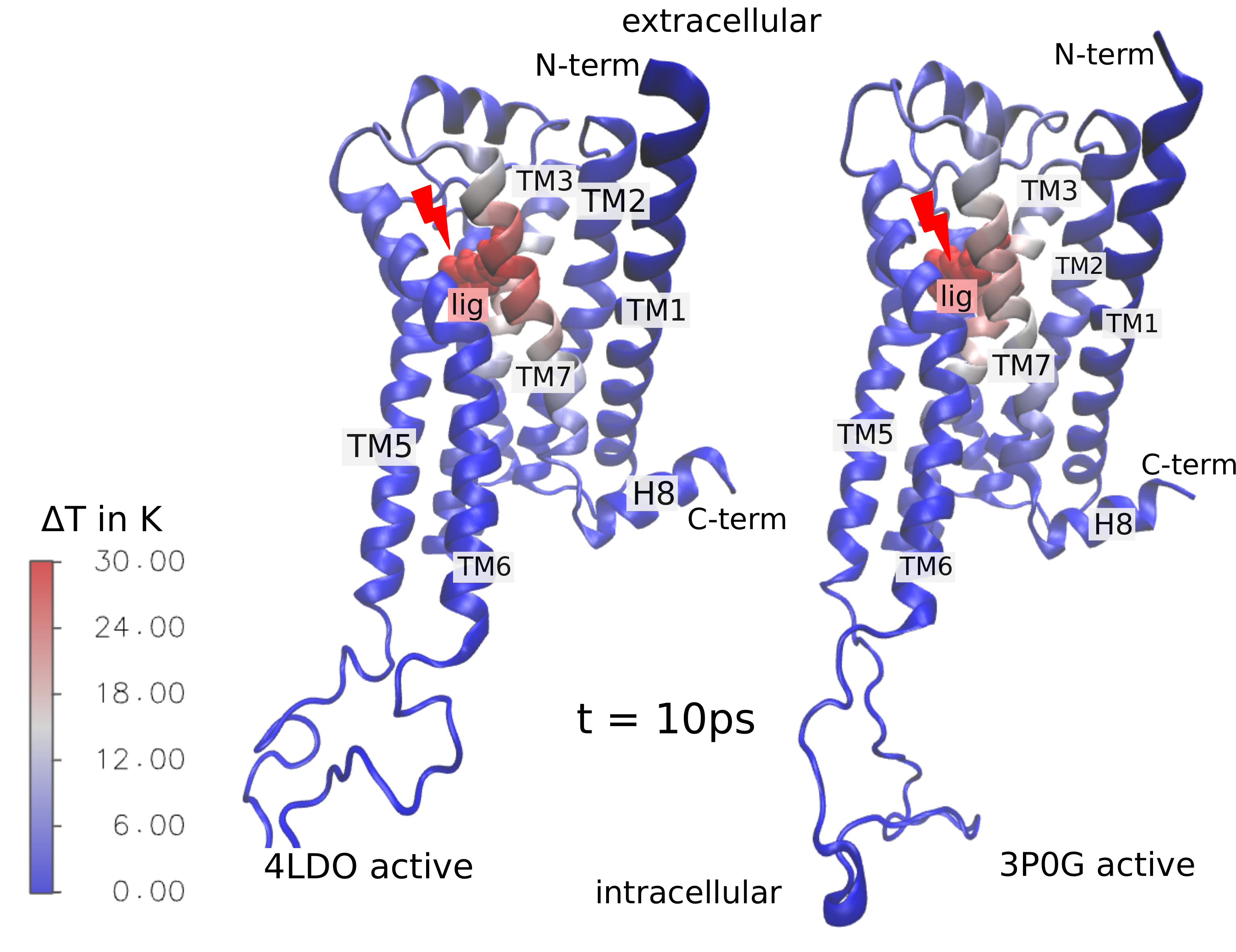}
	\caption{The two active states of $\beta_2AR$ (4LDO on the left and 3P0G on the right side) shown \SI{10}{ps} after heating the ligand.}
	\label{fig:B2AR_heaterLig}
\end{figure}

\begin{figure}[h!]
	\centering
	\includegraphics[width=.6\linewidth]{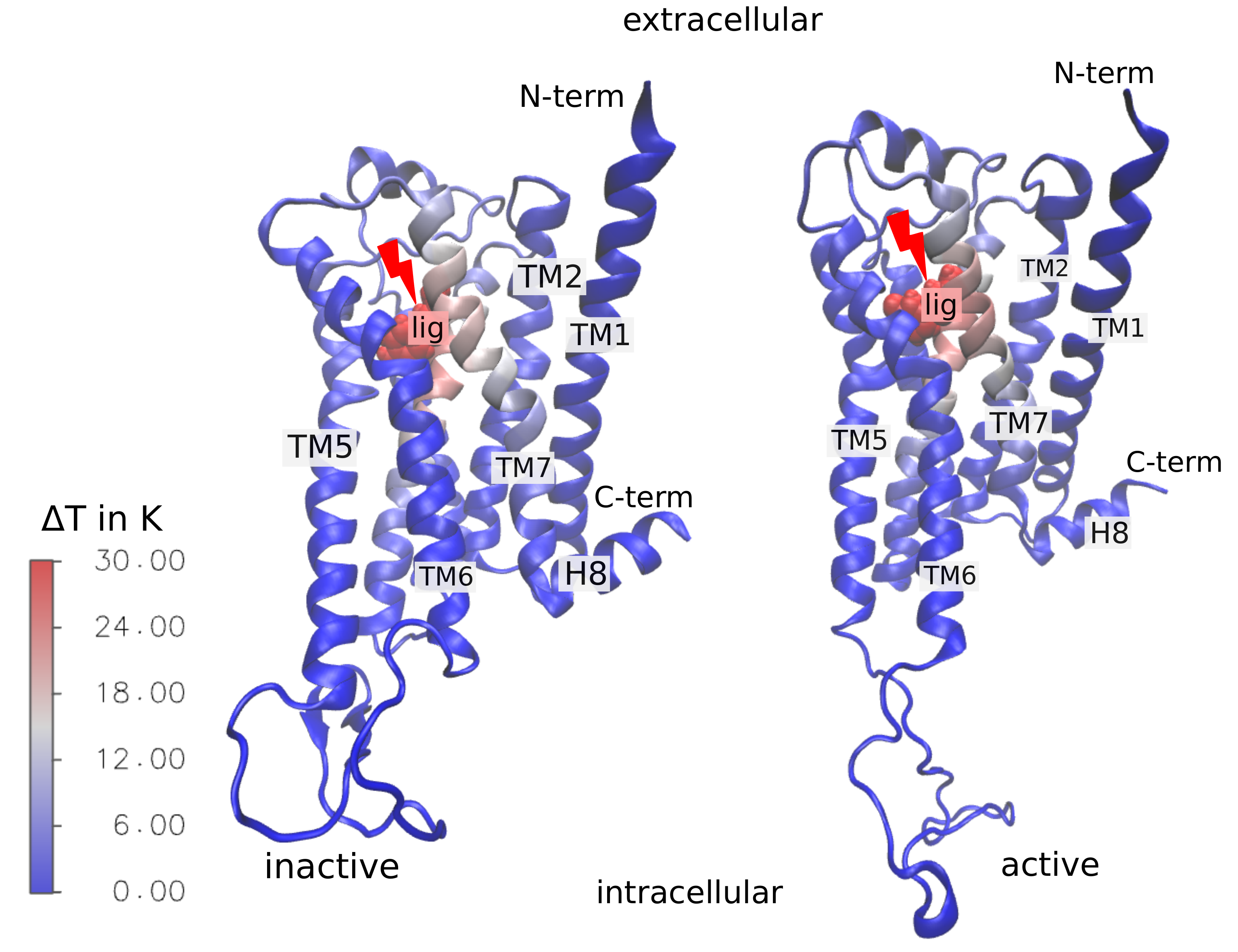}
	\caption{$\beta_2$AR shown in the inactive and active state \SI{10}{ps} after heating the ligand. The results are based on the contacts computed from 7 snapshots from the simulation trajectory.}
	\label{fig:B2AR_heaterLig_long}
\end{figure}